\IfFileExists{switch.tex}{\input{switch.tex}}{}
\makeatletter
\@ifundefined{fplong}{%
  \documentclass[nonatbib]{sigplanconf}
}{%
  \documentclass[nonatbib,nocopyrightspace]{sigplanconf}
}
\makeatother
\newenvironment{courte}{}{}
\newenvironment{longue}{\comment}{\endcomment}
\usepackage{amsmath}
\usepackage{amstext}
\usepackage{amssymb}
\usepackage{amsthm}
\usepackage[T1]{fontenc}
\usepackage[utf8x]{inputenc}
\usepackage{moreverb}
\usepackage{url}
\usepackage{graphicx}
\usepackage{listings}
\usepackage{xspace}
\usepackage{tabularx}
\usepackage{mathpartir}
  \renewcommand{\RefTirName}[1]{\hypertarget{#1}{\TirName {#1}}}

  \let\Rule\DefTirName
\usepackage{tikz}
\usepackage{paralist}
\input{macros} 
\usepackage{mymacros}

\newtheorem{lemma}{Lemma}
\newtheorem{definition}{Definition}

\begin{document}

\setlength{\pdfpageheight}{\paperheight}
\setlength{\pdfpagewidth}{\paperwidth}

\exclusivelicense
\conferenceinfo{ICFP~'13}{September 25--27, 2013, Boston, MA, USA}
\copyrightyear{2013}
\copyrightdata{978-1-4503-2326-0/13/09}
\doi{2500365.2500598}

\title{Programming with Permissions in \mezzo}

\authorinfo{François Pottier}
           {INRIA}
           {francois.pottier@inria.fr}
\authorinfo{Jonathan Protzenko}
           {INRIA}
           {jonathan.protzenko@ens-lyon.org}

\maketitle

\begin{abstract}
  We present \mezzo, a typed programming language of ML lineage. \mezzo is
  equipped with a novel static discipline of duplicable and affine
  permissions, which controls aliasing and ownership. This rules out certain
  mistakes, including representation exposure and data races, and enables new
  idioms, such as gradual initialization, memory re-use, and (type)state
  changes. Although the core static discipline disallows sharing a mutable
  data structure, \mezzo offers several ways of working around this
  restriction, including a novel dynamic ownership control mechanism which we
  dub ``adoption and abandon''.
\end{abstract}



\section{Introduction}
\label{sec:intro}

Programming with mutable, heap-allocated data structures is hard. In many
typed imperative programming languages, including Java, C\#, and ML, the type
system keeps track of the structure of objects, but not of how they are
aliased. As a result, a programming mistake can cause undesired sharing, which
in turn leads to breaches of abstraction, invariant violations, race
conditions, and so on. Furthermore, the fact that sharing is uncontrolled
implies that the type of an object must never change. This forbids certain
idioms, such as delayed initialization, and prevents the type system from
keeping track of the manner in which objects change state through method
calls. In order to work around this limitation, programmers typically use C\#
and Java's null pointer, or ML's option type. This implies that a failure to
follow an intended protocol is not detected at compile time, but leads to a
runtime error. In short, there is a price to pay for the simplicity of
traditional type systems: the bugs caused by undesired sharing, or by the
failure to follow an object protocol, are not statically detected.

This paper presents the design of a new programming language, \mezzo, which
attempts to address these issues. One motivating principle behind the design
of \mezzo is that one should be able to express precise assertions about the
current \emph{state} of an object or data structure. The type system should
keep track of \emph{state changes} and forbid using an object in a manner that
is inconsistent with its current state. An example is a socket that moves from
state ``ready'', to ``connected'', then to ``closed''. The ``close'' function,
for instance, should be invoked only if the socket is currently in the state
``connected'', and changes its state to ``closed''. Another example is a
collection, which must not be accessed while an iterator exists, but can be
used again once iteration is over.

Although state and state change play an important role in many programs, no
mainstream programming language builds these notions into its static
discipline. External tools must be used, such as typestate checking
tools~\cite{deline-faehndrich-04,bierhoff-aldrich-07,bierhoff-beckman-aldrich-09}
or tools for constructing proofs of programs, based for instance on separation
logic~\cite{smallfoot-05,verifast} or on the Spec\#
methodology~\cite{barnett-spec-04}.
Instead, we explore the possibility of reasoning about state within the type
system. This has well-known potential benefits. A property that is expressed
as a type is checked early, often, and at little cost. Furthermore, we believe
that, in the future, such a type system can serve as a strong foundation for
performing proofs of programs.

Obviously, if two ``principals'' separately think that ``the socket~$s$ is
currently connected'', and if one of them decides to close this socket, then
the other will be left with an incorrect belief about~$s$. Thus, precise
reasoning about state and state changes requires that information about a
mutable object (or data structure) be recorded in at most ``one place'' in the
type system. In \mezzo, this place is a \emph{permission}. Permissions keep
track not only of the structure of data, as does a traditional type system,
but also of must-alias and must-not-alias (i.e. equality and disjointness)
information. Like a separation logic assertion~\cite{reynolds-02}, a
permission has an ownership reading: to have access to a description of a part
of the heap is to own this part of the heap. Because ``to describe is to
own'', we need not explicitly annotate types with owners, as done in Ownership
Types~\cite{clarke-potter-noble-98} or Universe Types~\cite{dietl-mueller-05}.

We do not think of the ``type'' of an object and of its ``state'' as two
distinct notions: a permission describes both at once. Whereas previous work
on permissions~\cite{bierhoff-aldrich-07} distinguishes between a fixed type
structure and ``permissions'' that evolve with time, in \mezzo, both ``type''
and ``state'' can change over time. This yields greater expressiveness: for
instance, gradual initialization and memory re-use become possible. This also
yields greater simplicity and conciseness: for instance, when we write
polymorphic code that manipulates a list, a single type variable~$\tyvar$
denotes not only ``what'' the list elements are (e.g., sockets) but also in
what ``state'' they are and to what extent we ``own'' them.

The choices described above form our basic design premises. \mezzo can be
viewed as an experiment, whose aim is to determine to what extent these
choices are viable. Beyond these decisions, we strive to make the language as
simple as possible. \mezzo is a high-level programming language: we equip it
with first-class functions, algebraic data types, and require a garbage
collector. We could have chosen classes and objects instead of (or in addition
to) algebraic data types; this could be a topic for future research. We equip
\mezzo with a simple distinction between duplicable permissions (for immutable
data) and exclusive permissions (for mutable data). Although more advanced
varieties of permissions exist in the literature, including read-only views of
mutable data and fractional permissions~\cite{boyland-nesting-10}, we wish to
evaluate how far one can go without these advanced notions; if desired, they
could in principle be added to \mezzo.

By default, \mezzo's permission discipline imposes a restrictive aliasing
regime: the mutable part of the heap must form a forest. \mezzo offers several
mechanisms for evading this restriction. One, adoption and abandon, is new. It
allows arbitrary aliasing patterns within a region of the heap and achieves
soundness via dynamic checks. We describe it in detail in \sref{sec:adoption}.
The second mechanism is Boyland's nesting~\cite{boyland-nesting-10}. It can be
viewed as a form of adoption and abandon that requires no runtime checks but
is (for many purposes) less powerful. The last mechanism is locks in the
style of concurrent separation
logic~\cite{ohearn-07,gotsman-storable-07,hobor-oracle-08,buisse-11}.

\mezzo's static discipline has been formally defined and mechanically proved
sound\footnote{%
The formalization concerns a slightly lower-level language, Core Mezzo.
In Core Mezzo, fields are numbered, whereas in \mezzo they are named and
field names can be overloaded. At present, Core Mezzo is missing some of
the features of \mezzo, including parameterized
algebraic data types and mode constraints. We hope to add them in the future.}.
The formalization, which is available online~\cite{mezzo-proof}, includes
adoption and abandon, but does not (at present) cover nesting, locks, or
concurrency. The statement of soundness guarantees that ``well-typed programs
do not go wrong'', except possibly when the dynamic check performed at
``abandon'' fails.  In a concurrent extension of \mezzo, it would in
addition guarantee that ``well-typed programs are data-race-free''.
The proof is syntactic. We extend the typing judgement to a pair of a program
under execution and its heap, and establish the standard subject reduction and
progress lemmas.

A prototype type-checker has been implemented and is publicly
available~\cite{mezzo}. Several small libraries, totaling a few thousand lines
of code, have been written, and are also available online~\cite{mezzo}. They
include immutable data structures (lists), mutable data structures (lists,
doubly-linked lists, binary search trees, hash tables, resizable arrays, and
FIFO queues, see \sref{sec:adoption}), persistent data structures
implemented via imperative means (suspensions, persistent arrays), and a few
algorithms (memoization; graph search). At the time of this writing, an
interpreter is available, and a simple compiler (which translates \mezzo down
to untyped \ocaml) is being developed.

The paper begins with a motivating example (\sref{sec:example}), which cannot
be type-checked in ML, and which serves to informally illustrate \mezzo's
permission discipline. Then, we define the syntax of types, permissions, and
expressions (\sref{sec:syntax}) and informally explain the ownership reading
of permissions for immutable and mutable data (\sref{sec:ownership}). We
present the typing rules (\sref{sec:typing}) and introduce a few syntactic
conventions that make the surface language more palatable
(\sref{sec:surface-syntax}). We explain adoption and abandon, illustrate them
with a second example (\sref{sec:adoption}), and discuss nesting and locks
more briefly (\sref{sec:escape}).  Finally, we explain where \mezzo lies in
the design space and compare it with some of the previous approaches found in
the literature (\sref{sec:related}).

\begin{figure}[t]
  \begin{lstlisting}[numbers=left]
data list a =
   Nil |  Cons { head: a; tail: list a } -- \label{line:def-list}

data mutable mlist a =
  MNil | MCons { head: a; tail: list a } -- \label{line:def-mlist}

val rec appendAux [a] ( -- \label{line:aux}
    consumes dst: MCons { head: a; tail: () },
    consumes xs: list a,
    consumes ys: list a) : (| dst @ list a) = -- \label{line:post}
  match xs with
  | Nil -> -- \label{line:aux-nil}
      dst.tail <- ys;
      tag of dst <- Cons
  | Cons -> -- \label{line:aux-cons}
      let dst' = MCons { head = xs.head;
                         tail = () } in
      dst.tail <- dst';
      tag of dst <- Cons; -- \label{line:tagof}
      appendAux (dst', xs.tail, ys)
  end

val append [a] ( -- \label{line:append}
    consumes xs: list a,
    consumes ys: list a) : list a =
  match xs with
  | Nil -> -- \label{line:append-nil}
      ys
  | Cons -> -- \label{line:append-cons}
      let dst = MCons { head = xs.head; -- \label{line:append-dst}
                        tail = () } in
      appendAux (dst, xs.tail, ys); -- \label{line:append-aux}
      dst
  end
  \end{lstlisting}
  \caption{Tail-recursive concatenation of immutable lists}
  \label{fig:list}
\end{figure}

\section{\mezzo by example}
\label{sec:example}

\fref{fig:list} presents code for the concatenation of two immutable
lists. This example showcases several of \mezzo's features, and allows us to
explain the use of permissions. We review the code first
(\sref{sec:example:code}), then briefly explain how it is type-checked
(\sref{sec:example:permissions} and \sref{sec:example:loop}).

\subsection{Code}
\label{sec:example:code}

Our purpose is to write code that concatenates two \emph{immutable} lists
\li|xs| and \li|ys| to produce a new \emph{immutable} list. The traditional,
purely functional implementations of concatenation have linear space overhead,
as they implicitly or explicitly allocate a reversed copy of \li|xs|. Our
implementation, on the other hand, is written in \emph{destination-passing
  style}, and has constant space overhead. Roughly speaking, the list \li|xs|
is traversed and copied on the fly. When the end of \li|xs| is reached, the
last cell of the copy is made to point to \li|ys|.

The \li|append| function (\fref{fig:list}, line \ref{line:append}) is where
concatenation begins. If \li|xs| is empty, then the concatenation of \li|xs|
and \li|ys| is \li|ys| (\lineref{line:append-nil}). Otherwise (line
\ref{line:append-cons}), \li|append| allocates an \emph{unfinished, mutable}
cell \li|dst| (\lineref{line:append-dst}). This cell contains the first
element of the final list, namely \li|xs.head|. It is in an intermediate
state: it cannot be considered a valid list, since its \li|tail| field
contains the unit value~\li|()|. It is now up to \li|appendAux| to finish the
work by constructing the concatenation of \li|xs.tail| and \li|ys| and writing
the address of that list into \li|dst.tail|. Once \li|appendAux| returns,
\li|dst| \emph{has become} a well-formed list (this is indicated by the
postcondition \qli|dst @ list a| on \lineref{line:post}) and is returned by \li|append|.

The function \li|appendAux| expects an unfinished, mutable cell~\li|dst|
and two lists \li|xs| and \li|ys|. Its purpose is to write the
concatenation of \li|xs| and \li|ys| into \li|dst.tail|, at which point
\li|dst| can be considered a well-formed list.
If \li|xs| is \li|Nil| (\lineref{line:aux-nil}), the \li|tail| field of
\li|dst| is made to point to \li|ys|. Then, \li|dst|, a mutable \li|MCons|
cell, is ``frozen'' by a tag update instruction and becomes an immutable
\li|Cons| cell. (This instruction compiles to a no-op.) If \li|xs| is a
\li|Cons| cell (\lineref{line:aux-cons}), we allocate a new destination cell
\li|dst'|, let \li|dst.tail| point to it, freeze \li|dst|, and repeat the
process via a tail-recursive call.

This example illustrates several important aspects of \mezzo.

\paragraph{Expressiveness} In a traditional typed programming language, such as
  Java or \ocaml, list concatenation in destination-passing style is possible,
  but its result must be a mutable list, because an \emph{immutable} list cell
  cannot be gradually initialized.

\paragraph{State change} The call \li|appendAux(dst, xs, ys)| changes the ``type''
  of \li|dst| from ``unfinished, mutable list cell'' to ``well-formed,
  immutable list''. This type-changing update is sound because one must be the
  ``unique owner'' of the mutable cell \li|dst| for this call to be permitted.

\paragraph{Ownership transfer} In fact, the call \li|appendAux(dst, xs, ys)| also
  changes the ``type'' of \li|xs| and \li|ys| from ``immutable list'' to
  ``unknown''. Indeed, the postcondition of \li|appendAux| guarantees nothing
  about \li|xs| and \li|ys|. In other words, the caller gives up the
  permission to use \li|xs| and \li|ys| as lists, and in return gains the
  permission to use \li|dst| as a list. In other words, the ownership of the list elements is
  transferred from \li|xs| and \li|ys| to \li|dst|. This is
  required for soundness. We do not know what the list elements are (they
  have abstract type~\li|a|). They could be mutable
  objects, whose ``unique ownership'' property must not be violated%
  \footnote{We later note (\sref{sec:modes}) that if at a call site the variable \li|a| is
    instantiated with a duplicable type, say \li|int|, then the
    permissions \li|xs @ list int| and \li|ys @ list int| are
    considered duplicable, so they can in fact be duplicated prior to the call
    \li|appendAux(dst, xs, ys)|, hence are not lost.
}.

\subsection{Permissions}
\label{sec:example:permissions}

Permissions do not exist at runtime: they are purely an artefact of the type
system. An atomic permission $\tyatomic{x}{t}$ represents
the right to use the program variable $x$ at type~$t$. Two permissions $P_1$
and~$P_2$ can be combined to form a composite permission $\tystar{P_1}{P_2}$. The
conjunction $\ast$ is separating~\cite{reynolds-02} at mutable memory
locations and requires agreement at immutable locations (\sref{sec:modes}).
The empty permission, a unit for conjunction, is written $\tyempty$.

When execution begins, a program conceptually possesses an empty
permission. As execution progresses through the code, permissions come and
go. At any program point, there is a certain \emph{current permission}. Most
of the time, the manner in which permissions evolve and flow is implicit. It
must be made explicit in a few places: in particular, every function type
must include explicit pre- and postconditions.

Let us continue our discussion of the concatenation example
(\fref{fig:list}). We explain in an informal manner how the
function \li|append| is type-checked. This allows us to illustrate
how permissions are used and how they evolve.

The typing rules appear in \fref{fig:typing-rules};
\begin{courte}
a subset of
\end{courte}
the permission subsumption rules appear in \fref{fig:sub}. In the following,
we refer to some of these rules, but defer their detailed explanation to
\sref{sec:typing}.

The \li|append| function is defined at \lineref{line:append}. At the beginning
of the function's body, by the typing rule \Rule{Function}, permissions for
the formal arguments are available. Thus, the current permission is:
$$
\tystar%
  {\tyatomic{xs}{\tylist{a}}}%
  {\tyatomic{ys}{\tylist{a}}}%
$$
This permission represents the right to use \li|xs| and \li|ys| as lists of
elements of type $a$.

This permission soon evolves, thanks to the \li|match| construct, which
examines the tag carried by \li|xs|. By the typing rule \Rule{Match}, as we learn that \li|xs| is a
\li|Nil| cell, we replace our permission about \li|xs| with a more precise
one, which incorporates the knowledge that the tag of \li|xs| is \li|Nil|.
At \lineref{line:append-nil}, the current permission becomes:
$$
\tystar%
  {\tyatomic{xs}{\kw{Nil}}}%
  {\tyatomic{ys}{\tylist{a}}}%
$$
$\tyatomic{xs}{\kw{Nil}}$ is a \emph{structural permission}: it
asserts that \li|xs| points to a memory block whose tag is $\kw{Nil}$ (and
which has zero fields). Similarly, at \lineref{line:append-cons}, the current
permission becomes:
$$
\tystar%
  {\tyatomic{xs}{\tyconcrete{Cons}{\fhead: a; \ftail: \tylist{a}}}}%
  {\tyatomic{ys}{\tylist{a}}}%
$$
The structural permission for \li|xs| asserts that \li|xs| points to a memory
block that carries the tag \kw{Cons} and has a \li|head| field of type $a$ and
a \li|tail| field of type $\tylist{a}$.

At this stage, the type-checker performs an implicit operation. It applies the
permission subsumption rule \Rule{DecomposeBlock}. This causes fresh names
$hd$ and $tl$ to be introduced for the \li|head| and \li|tail| fields of this
structural permission. This yields the following conjunction:
\[\begin{array}{r@{}l}
\tyatomic{xs&}{\tyconcrete{Cons}{\fhead: (\tysingleton{hd}); \ftail: (\tysingleton{tl})}}
\ast{} \\
\tyatomic{hd&}{a}
\ast
\tyatomic{tl}{\tylist{a}}
\ast{} \\
\tyatomic{ys&}{\tylist{a}}
\end{array}\]
This is our first encounter of a singleton type, which we write~$\tysingleton{hd}$. A permission of the form
$\tyatomic{x}{\tysingleton{y}}$ asserts that the variables~$x$ and~$y$ denote
the same value. In particular, if they denote memory locations, this
means that $x$ and $y$ point to the same object:
this is a \emph{must-alias}
constraint. We write $x = y$ for $\tyatomic{x}{\tysingleton{y}}$.
Similarly, in the structural permission above, the fact that the $\fhead$
field has
type $\tysingleton{hd}$ means that the value of this field is~$hd$. We write
$\fhead = hd$ for $\fhead: (\tysingleton{hd})$.

By the typing rules \Rule{New} and \Rule{Let}, when the cell \li|dst| is
allocated (\lineref{line:append-dst}), a permission for \li|dst| appears,
namely:
\[\tyatomic{dst}{%
  \tyconcrete{MCons}{ \fhead = hd; \ftail: () } }\]
We now see how singleton types help reason about sharing. At this point, we
have three permissions that mention $hd$.  We know that $hd$ is stored in the
$\fhead$ field of $xs$; we know that $hd$ is stored in the $\fhead$ field of
$dst$; and we have a permission to use $hd$ at type $a$. We do not need a
borrowing convention~\cite{aldrich-borrowing-12} in order to fix which of $xs$
or $dst$ owns $hd$. Instead, the system knows that the object $hd$ is accessible
via two paths, namely $xs.\fhead$ and $dst.\fhead$, and can be used under
either name. This use of singleton types is taken from Alias Types~\cite{alias-types-00}.

By the typing rules \Rule{Read} and \Rule{Application}, in order to call
\li|appendAux(dst, xs.tail, ys)| (\lineref{line:append-aux}), we need the
following conjunction of permissions. It is the precondition of
\li|appendAux|, suitably instantiated:
$$
\tystar
{\tyatomic{dst}{\tyconcrete{MCons}{ \fhead: a; \ftail: () }}}
\tystar
{\tyatomic{tl}{\tylist{a}}}
{\tyatomic{ys}{\tylist{a}}}
$$
Are we able to satisfy this requirement? The answer is positive. The
subsumption rules \Rule{ExistsIntro} and \Rule{DecomposeBlock} allow combining the permissions
$\tyconcrete{MCons}{ \fhead = hd; \ftail: () }$ and
$\tyatomic{hd}{a}$ (both of which are present) to obtain the first
conjunct above. The second and third conjuncts above are present already.

By \Rule{Application}, the precondition of \li|appendAux| is consumed (taken
away from the caller). After the call, the postcondition of \li|appendAux|
is added to the current permission, which is then:
\[
\tyatomic{xs}{\tyconcrete{Cons}{\fhead = hd; \ftail = tl}}
\ast
\tyatomic{dst}{\tylist{a}}
\]
The conjunct that concerns $xs$ is of no use, and is in fact silently
discarded when we reach the end of the \kw{Cons} branch within \li|append|.
The conjunct that concerns $dst$ is used to check that this branch satisfies
\li|append|'s advertised return type, namely
$\tylist{a}$. Similarly, in the $\kw{Nil}$ branch,
the permission $\tyatomic{ys}{\tylist{a}}$ shows
that a value of appropriate type is returned.
In conclusion, \li|append| is well-typed.

\subsection{To loop or to tail call?}
\label{sec:example:loop}

In-place concatenation (that is, melding) of mutable lists can also be
implemented by a tail-recursive function. The pattern is analogous to that
of \fref{fig:list}, but the code is simpler, because the first list is not
copied, and ``freezing'' is not required.

These algorithms are traditionally viewed as iterative and implemented using
a \li|while| loop. Berdine \etal.'s iterative formulation of mutable list
melding~\cite{smallfoot-05}, which is proved correct in separation logic, has
a complex loop invariant, involving two ``list segments'', and requires an
inductive proof that the concatenation of two list segments is a list segment.
In contrast, in the tail-recursive approach, the ``loop invariant'' is the
type of the recursive function (e.g., \li|appendAux| in \fref{fig:list}).
This type is reasonably natural and does not involve list segments.

How do we get away without list segments and without inductive reasoning?
The trick is that, even though \li|appendAux| is tail-recursive, which means
that no code is executed after the call by \li|appendAux| to itself, a
\emph{reasoning step} still takes place after the call. Immediately before
the call, the current permission can be written as follows:
\newcommand{\permcons}[3]{\tyatomic{#1}{\tyconcrete{Cons}{\fhead#2; \ftail#3}}}
\newcommand{\permmcons}[3]{\tyatomic{#1}{\tyconcrete{MCons}{\fhead#2; \ftail#3}}}
\newcommand{\astnl}{\ast{}\\}
\[\begin{array}{r@{}l}
\permcons{xs&}{=hd}{=tl}
\astnl
\permcons{dst&}{: a}{=dst'}
\astnl
\permmcons{dst'&}{: a}{: ()}
\astnl
\tyatomic{tl&}{\tylist{a}} \ast \tyatomic{ys}{\tylist{a}}
\end{array}\]
The call \qli|appendAux (dst', xs.tail, ys)| consumes the last three
permissions and produces instead $\tyatomic{dst'}{\tylist{a}}$.
The first two permissions
are ``framed out'', i.e., implicitly preserved. After the call, we have:
\[\begin{array}{r@{}l}
\permcons{xs&}{=hd}{=tl}
\astnl
\permcons{dst&}{: a}{=dst'}
\astnl
\tyatomic{dst'&}{\tylist{a}}
\end{array}\]
Dropping the first permission and combining the last two yields:
\[\begin{array}{r@{}l}
\permcons{dst&}{: a}{: \tylist{a}}
\end{array}\]
which can be folded back to $\tyatomic{dst}{\tylist{a}}$,
so \li|appendAux| satisfies its postcondition.
\begin{longue}
The framing out of a permission during the recursive call, as well as the
folding step that takes place after the call, are the key technical mechanisms
that allow us to avoid the need for list segments and inductive reasoning.
\end{longue}
In short, the code is tail-recursive, but the manner in
which one reasons about it is recursive.

Minamide~\cite{minamide-98} proposes a notion of ``data structure with a
hole'', or in other words, a segment, and applies it to the problem of
concatenating immutable lists.
Walker and Morrisett~\citeyear{recursive-alias-types-00} offer a
tail-recursive version of mutable list concatenation in
a low-level typed intermediate language, as opposed to a surface language.
The manner in which they avoid reasoning about list segments is analogous to
ours. There, because the code is formulated in continuation-passing style, the
reasoning step that takes place ``after the recursive call'' amounts to
composing the current continuation with a coercion.
Maeda \etal.~\citeyear{maeda-11} study a slightly different approach, also in
the setting of a typed intermediate language, where separating implication
offers a way of defining list segments.
Our approach could be adapted to an iterative setting by adopting a new
proof rule for \li|while| loops. This is noted independently by
Charguéraud~\cite[\S3.3.2]{chargueraud-10} and by Tuerk~\citeyear{tuerk-10}.

\section{Syntax}
\label{sec:syntax}

  \begin{figure}[t]
    \begin{center}
      \begin{tabularx}{\columnwidth}{rX}
        $\kappa ::=$
        & $\ktype \mid \kterm \mid \kperm \mid \kappa\rightarrow\kappa$ \hfill kind \\
          &\\

        $T, t, P ::=$ & \hfill type or permission \\
          & $X                                $ \hfill variable ($\tyvar, x, \ldots$) \\
          & $\tyarrow{t}{t}                   $ \hfill function type \\
          & $\tytuple{\vec{t}\,}              $ \hfill tuple type \\
          & $\tyconcreteadopts{A}{\vec{f}: \vec{t}}{t} $ \hfill structural type \\
          & $\tyapp{T}{\vec{T}}               $ \hfill $n$-ary type application \\
          & $\tyforall{X}{\kappa}{T}          $ \hfill universal quantification \\
          & $\tyexists{X}{\kappa}{T}          $ \hfill existential quantification \\
          & $\tysingleton{x}                  $ \hfill singleton type \\
          & $\tybar{t}{P}                     $ \hfill type/permission conjunction \\
          & $\tydynamic                       $ \hfill (see \sref{sec:adoption}) \\
          & $\tyatomic{x}{t}      $ \hfill atomic permission \\
          & $\tyempty                         $ \hfill empty permission \\
          & $\tystar{P}{P}                    $ \hfill permission conjunction \\
          & $\tynameintro{x}{t}               $ \hfill \underline{name introduction} (see \sref{sec:surface-syntax}) \\
          & $\tyconsumes{T}                   $ \hfill \underline{consumes annotation} (see \sref{sec:surface-syntax}) \\
          &\\

        $d ::= $ & \hfill algebraic data type definition \\
          & $\kw{mutable}?\; \kw{data } d\ (\vec{X}: \vec\kappa) =
          \,\vec{b}$ \\ & $ 
          \quad\kw{adopts}\,t$ \\
         &\\

         $b ::= $ & $\tyconcrete{A}{\vec{f}: \vec{t}} $ \hfill algebraic data type branch \\

      \end{tabularx}

    \end{center}
    \caption{Syntax of types and permissions}
    \label{fig:mezzo-types}
  \end{figure}

  \begin{figure}[t]
    \begin{center}
      \begin{tabularx}{\columnwidth}{rX}
        $e ::= $ & \hfill expression \\
          & $x                           $ \hfill variable \\
          & $\elet{p}{e}{e}              $ \hfill local definition \\
          & $\efun{\tyforalln{\vec{X}: \vec\kappa}{(x:t)}}{t}{e}$ \hfill anonymous function \\
          & $\etapply{e}{t: \kappa}      $ \hfill type instantiation \\
          & $\eapply{e}{e}               $ \hfill function application \\
          & $\etuple{\vec{e}\,}          $ \hfill tuple \\
          & $\econstruct{A}{\vec{f} = \vec{e}} $ \hfill data constructor application \\
          & $\eaccess{e}{f}              $ \hfill field access \\
          & $\eassign{e}{f}{e}           $ \hfill field update \\
          & $\ematch{e}{\ \vec{p} \rightarrow \vec{e}} $ \hfill case analysis \\
          & $\eassigntag{e}{A}           $ \hfill tag update \\
          & $\egive{e}{e}                $ \hfill adoption \\
          & $\etake{e}{e}                $ \hfill abandon \\
          & $\efail                      $ \hfill dynamic failure \\
         &\\

        $p ::= $ & \hfill pattern \\
          & $x                           $ \hfill variable \\
          & $\etuple{\vec{p}\,}          $ \hfill tuple pattern \\
          & $\econstruct{A}{ \vec{f} = \vec{p} } $ \hfill data constructor pattern \\

      \end{tabularx}
    \end{center}
    \caption{Syntax of expressions}
    \label{fig:mezzo-expressions}
  \end{figure}

\subsection{Types}
\label{sec:syntax-types}

We work with the ``internal syntax'' of types. The surface syntax adds a few
syntactic conventions, which we explain later on
(\sref{sec:surface-syntax}). For the moment, the reader may ignore the
two underlined constructs in \fref{fig:mezzo-types}.

Types have kinds. The base kinds are $\ktype$, $\kterm$, and $\kperm$. The
standard types, such as function types, tuple types, etc.\ have kind $\ktype$.
The types of kind $\kterm$ are program variables. If a variable~$x$ is bound
(by \kw{let}, \kw{fun}, or \kw{match}) in the code, then $x$ may appear not
only in the code, but also in a type: it is a type of kind~$\kterm$. The types
of kind $\kperm$ are permissions. First-order arrow kinds are used to classify
parameterized algebraic data types.

In \fref{fig:mezzo-types}, we use the meta-variables $T$ and $X$ to stand for
types and variables of arbitrary kind; we use $t$ and $P$ to suggest that a
type has kind $\ktype$ and $\kperm$, respectively; we use $\tyvar$ and~$x$ to
suggest that a variable has kind $\ktype$ and $\kterm$, respectively.
\begin{courte}
We omit the definition of the kinding judgment; it appears in the extended
version of this paper~\cite{self-long}.
\end{courte}
\begin{longue}
The full definition of the kind system appears in Appendix~\ref{sec:kinding-appendix}.
\end{longue}

The \emph{structural type} $\tyconcrete{A}{\vec{f}: \vec{t}}$
describes a block in the heap whose tag is currently $\kw{A}$ and whose fields
$\vec{f}$ currently have the types~$\vec{t}$. An example, taken from
\sref{sec:example}, is $\tyconcrete{MCons}{
    \fhead: a; \ftail: () }$. The data constructor \kw{A} must refer to a previously defined
algebraic data type, and the fields $\vec{f}$ must match the definition of~\kw{A}.
The types~$\vec{t}$, however, need not match the types that appear in the
definition of~$\kw{A}$. For instance, in the definition of \kw{MCons}, the
type of the $\ftail$ field is $\tymlist{a}$, not $()$. This implies that
the above structural type cannot be folded to $\tymlist{a}$; the $\ftail$
field must be updated first.
A structural type may include a clause of the form $\kw{adopts}\,t$, whose
meaning is explained later on (\sref{sec:adoption}). If omitted,
$\kw{adopts}\,\bot$ is the default.

An example of a type application $\tyapp{T}{\vec{T}}$ is $\tylist\tyint$.  We
sometimes refer to this as a \emph{nominal type}, as opposed to a structural
type.

The universal and existential types are in the style of
System~$F$. A (base) kind annotation is mandatory; if omitted, $\ktype$ is the
default. The bottom type $\bot$ and the top type $\tyunknown$ can be defined
as $\forall\tyvar.\tyvar$ and $\exists\tyvar.\tyvar$, respectively.

The conjunction of a type and a permission is written $\tybar{t}{P}$. Because
permissions do not exist at runtime, a value of this type is represented at
runtime as a value of type $t$. Such a conjunction is typically used to
express function pre- and postconditions.  The type $\tybar{()}{P}$ is
abbreviated as $\tybar{}{P}$.

Algebraic data type definitions are prefixed with the keyword $\kw{data}$. They
are anologous to Haskell's and \ocaml's. Each branch is explicitly
named by a data constructor and carries a number of named fields. If a
definition begins with the keyword $\kw{mutable}$, then the tag and all
fields are considered mutable, and can be modified via tag update and field
update instructions; otherwise, they are considered immutable. Examples appear
at the top of \fref{fig:list}. Like a structural type, an algebraic data type
definition may include an \kw{adopts} clause; if omitted, $\kw{adopts}\,\bot$
is the default.

\subsection{Expressions}

Expressions (\fref{fig:mezzo-expressions}) form a fairly
standard $\lambda$-calculus with tuples and algebraic data structures.
A function definition must be explicitly annotated with the function's type
parameters, argument type, and return type. One reason for this is that the
argument and return type serve as pre- and postconditions and in general
cannot be inferred.  Furthermore, we have System~$F$-style
polymorphism. Explicit type abstractions are built into function
definitions. Type applications must in principle be explicit as well. The
prototype type-checker allows omitting them and performs a limited form of
type inference, which is outside the scope of this paper.

\section{Ownership, modes, and extent}
\label{sec:ownership}

We wrote earlier (\sref{sec:intro}) that ``to have a permission for~$x$'' can
be understood informally as ``to own $x$''. Roughly speaking, this is true,
but we must be more precise, for two reasons. First, we wish to distinguish
between mutable data, on which we impose a ``unique owner'' policy, and
immutable data, for which there is no such restriction. For this reason, types
and permissions come in several flavors, which we refer to as \emph{modes} (\sref{sec:modes}).
Second, in a permission of the form $\tyatomic{x}{t}$, the
type~$t$ describes the \emph{extent} to which we own~$x$. If $xs$
is a list cell, do we own just this cell? the entire spine? the
spine and the elements? The answer is given by the type~$t$. For
instance (\sref{sec:extent}), $\tyatomic{xs}{\tyconcrete{Cons}{\fhead =
    hd; \ftail = tl}}$ represents the ownership of just the cell $xs$, because
the singleton types $\tysingleton{hd}$ and $\tysingleton{tl}$ denote the
ownership of an empty heap fragment. On the other hand,
$\tyatomic{xs}{\tyconcrete{Cons}{\fhead: a; \ftail:
    \tylist{a}}}$ gives access to the entire list spine. (Because \kw{list}
is an immutable algebraic data type, this is read-only, shared access.)  It
further gives access to all of the list elements, insofar as the type~$a$
allows this access. In this example, $a$ is a variable: one must wait
until $a$ is instantiated to determine what the elements are and to what
extent we own them.

\subsection{Modes}
\label{sec:modes}

A subset of the permissions are considered \emph{duplicable}, which means that
they can be implicitly copied (\Rule{Duplicate}, \fref{fig:sub}). Copying a
permission for an object~$x$ means that~$x$ may be shared: it may be
used via different pointers, or by different threads simultaneously. Thus, a
duplicable permission does not represent unique ownership; instead, it denotes
\emph{shared knowledge}. Because the system does not control with whom this knowledge is
shared, this knowledge must never be invalidated, lest some principals be
left with an outdated version of the permission.  Therefore, a duplicable
permission denotes \emph{shared, permanent knowledge}. The permissions that
describe read-only, immutable data are duplicable: for instance,
$\tyatomic{xs}{\tyconcrete{Cons}{\fhead = hd;\ftail=tl}}$ and
$\tyatomic{xs}{\tylist\tyint}$ are duplicable.

A subset of the permissions are considered \emph{exclusive}. An exclusive
permission for an object~$x$ represents the ``unique ownership'' of~$x$.
In other words, such a permission grants \emph{read-write access} to the
memory block at address~$x$ and guarantees that no-one else has access to
this block. The permissions that describe mutable memory blocks are
exclusive: for instance,
$\tyatomic{xs}{\tyconcrete{MCons}{\fhead = hd;\ftail=tl}}$ is exclusive.
An exclusive permission is analogous to a ``unique'' permission in other
systems~\cite{bierhoff-aldrich-07} and to a separation logic
assertion~\cite{reynolds-02}.

\begin{courte}
By lack of space, we must unfortunately omit the definition of the
predicates ``$t$ is duplicable'' and ``$t$ is exclusive'', which
are used in the typing rules (\fref{fig:typing-rules}). They can
be found in the extended version of this paper.  
\end{courte}
\begin{longue}
Predicates of the form ``$t$ is duplicable'' and ``$t$ is exclusive'' are part
of a more general form of predicates which we call \emph{facts}. We are able to
compute the fact for any given type, as well as an optimal fact for any given
data type, such as ``a list is duplicable as long as its elements are
duplicable''. The details are provided in \sref{sec:modes-appendix}.
\end{longue}

No permission is duplicable and exclusive. Some permissions are neither
duplicable nor exclusive. ``$\tyatomic{xs}{\tylist{(\tyref\tyint)}}$'', which describes an
immutable list of references to integers, illustrates this. It must not be
duplicated: this would violate the ``unique owner'' property of the list
elements. It is not exclusive: the list cell at~\li|xs| is an immutable
object, and this permission does not guarantee exclusive access to this cell.
Another example is \qli|x @ a|. Because \li|a| is a type variable, one cannot
assume that this permission is duplicable (or exclusive)\footnote{\mezzo allows
the programmer to explicitly assume that a type variable \li|a| is duplicable,
or exclusive. This mechanism is not treated in this paper.}.

Every permission is affine. One can implicitly drop a permission that one does
not need.

The language is designed so that the type-checker (and the programmer!) can
always tell what \emph{mode} a permission~$P$ satisfies: duplicable,
exclusive, or neither (hence, affine). Modes form an upper semi-lattice,
whose top element is ``affine'', and where ``duplicable'' and ``exclusive''
are incomparable.
Because algebraic data types are recursively defined, their mode analysis
requires a fixed point computation,
\begin{courte}
whose details we omit.
\end{courte}
\begin{longue}
whose details are given in \sref{sec:modes-appendix}.
\end{longue}

If $t$ and $u$ are exclusive types, then the conjunction
$\tystar{\tyatomic{x}{t}}{\tyatomic{y}{u}}$
implies that $x$ and $y$ are \emph{distinct} addresses. In other words,
\emph{conjunction of exclusive permissions is separating}.
On the other hand, if $t$ and/or $u$ are duplicable, $x$ and $y$ may be
aliases. Conjunction is not in general separating. \emph{Conjunction of
duplicable permissions} requires agreement between the two conjuncts.
The reader is referred to the draft paper that accompanies the type soundness
proof~\cite{mezzo-proof} for a formal definition of the semantics of conjunction.

\subsection{Extent}
\label{sec:extent}

Every type $t$ has an ownership reading: that is, the permission
$\tyatomic{x}{t}$ represents certain access rights
about $x$. However, the extent of these rights (or, in separation
logic terminology, their footprint) depends on the type~$t$.

A singleton type $\tysingleton{y}$, for instance, has empty extent. Indeed,
the permission $\tyatomic{x}{\tysingleton{y}}$, which we usually
write $x = y$, asserts that $x$ and $y$ are equal, but does not allow assuming
that $x$ is a pointer, let alone dereferencing it.

A structural type such as $\tyconcrete{\kw{Cons}}{\fhead = hd;
  \ftail = tl}$ has an extent of one memory block. The permission
$\tyatomic{xs}{\tyconcrete{\kw{Cons}}{\fhead = hd; \ftail =
    tl}}$ gives us (read-only, shared) access to the block at address~$xs$, and
guarantees that its $\fhead$ and $\ftail$ fields contain the values~$hd$
and~$tl$, respectively, but (as per the semantics of singleton types)
guarantees nothing about $hd$ and $tl$.

What is the extent of a ``deep'' composite type, such as the structural type
$\tyconcrete{\kw{Cons}}{\fhead: a; \ftail: \tylist{a}}$ or the nominal
type $\tylist{a}$? What does it mean to own a list? In order to answer these
questions, one must understand how a composite permission is decomposed into a
conjunction of more elementary permissions.

A structural permission, such as
$\tyatomic{xs}{\tyconcrete{\kw{Cons}}{\fhead: a; \ftail:
    \tylist{a}}}$, can be decomposed by introducing a fresh name for each of
the values stored in the fields.  (See \Rule{DecomposeBlock}
in \fref{fig:sub}.) The result is a more verbose, but
logically equivalent, permission:
$$
\exists hd, tl.(
\tystar
{\tyatomic{xs}{\tyconcrete{\kw{Cons}}{\fhead = hd; \ftail = tl}}}
\tystar
{\tyatomic{hd}{a}}
{\tyatomic{tl}{\tylist{a}}}
)
$$
The meaning and extent of the original structural permission is now clearer:
it grants access to the cell at~$xs$ \emph{and} to the first list
element (to the extent dictated by the type~$a$) \emph{and} to the rest
of the list.

The meaning of a nominal permission, such as
$\tyatomic{xs}{\tylist{a}}$, is
just the disjunction of the meanings of its
unfoldings, namely $\tyatomic{xs}{\tyconcretezero{Nil}}$
and $\tyatomic{xs}{\tyconcrete{\kw{Cons}}{\fhead: a; \ftail:
    \tylist{a}}}$.

If $a$ is (instantiated with) an exclusive type, then we find that
$\tyatomic{xs}{\tylist{a}}$ implies that the list elements
are \emph{pairwise distinct}, and grants read-only access to the list
spine and exclusive access to the list elements.

\begin{figure*}
  \begin{mathpar}
    \inferrule[Var]
    {}
    {K; \tyatomic{x}{t} \vdash x: t}

    \inferrule[Let]
    {K; P \vdash e_1: t_1 \\
    K, x: \kterm; \tyatomic{x}{t_1} \vdash e_2: t_2}
    {K; P \vdash \elet{x}{e_1}{e_2}: t_2}

    \inferrule[Function]
    {K, \vec{X}:\vec\kappa, x: \kterm; P \ast \tyatomic{x}{t_1} \vdash e : t_2\\
    P\text{ is duplicable}}
    {K; P \vdash \efun{[\vec X:\vec\kappa]\;(x: t_1)}{t_2}{e} :
     \tyforall{\vec{X}}{\vec\kappa}{t_1 \rightarrow t_2}}

    \inferrule[Instantiation]
    {K; P \vdash e : \tyforall{X}{\kappa}{t_1}}
    {K; P \vdash e : [T_2/X]t_1}

    \inferrule[Application]
    {}
    {K; \tyatomic{x_1}{t_2 \rightarrow t_1} \ast \tyatomic{x_2}{t_2} \vdash \eapply{x_1}{x_2}: t_1}

    \inferrule[Tuple]
    {}
    {K; \tyatomic{\vec{x}}{\vec{t}} \vdash (\vec{x}): (\vec{t}\,)}

    \inferrule[New]
    {\tyconcrete{A}{\vec{f}} \text{ is defined}}
    {K; \tyatomic{\vec{x}}{\vec{t}} \vdash
      \tyconcrete{A}{\vec{f} = \vec{x}}:
      \tyconcrete{A}{\vec{f}: \vec{t}}}

    \inferrule[Read]
    {t \text{ is duplicable} \\\\
     P \text{ is } \tyatomic{x}{
                     \tyconcreteadopts{A}{F[f: t]}{u}}
    }
    {K; P \vdash x.f: \tybar{t}{P}}

    \inferrule[Write]
    {\tyconcrete{A}{\ldots} \text{ is exclusive}}
    {
     \begin{array}{r@{\;}l}
     K; & \tyatomic{x_1}{\tyconcreteadopts{A}{F[f: t_1]}{u}} \ast
        \tyatomic{x_2}{t_2}
     \vdash
     x_1.f \leftarrow x_2: \tybar{}{ \cr&
        \tyatomic{x_1}{\tyconcreteadopts{A}{F[f: t_2]}{u}}
      }
     \end{array}
    }

    \inferrule[Match]{
    \text{for every } i, \quad
    K; P \vdash \elet{p_i}{x}{e_i} : t
    }
    {
      K; P \vdash \ematch{x}{\ \vec{p} \rightarrow \vec{e}} : t
    }

    \inferrule[WriteTag]
    {
    \tyconcrete{A}{\ldots} \text{ is exclusive}\\
    \tyconcrete{B}{\vec{f'}} \text{ is defined}\\
    \# \vec{f} = \# \vec{f'}
    }
    {
      \begin{array}{r@{\;}l}
      K; & \tyatomic{x}{
      \tyconcreteadopts{A}{\vec{f}: \vec{t}}{u}}
       \vdash
       \eassigntag{x}{B}: \tybar{}{\cr&
        \tyatomic{x}{
          \tyconcreteadopts{B}{\vec{f'}: \vec{t}}{u}}
      }
      \end{array}
    }

    \inferrule[Give]{
    t_2 \text{ adopts } t_1
    }{
      \begin{array}{r@{\;}l}
      K;
      \tyatomic{x_1}{t_1} \ast&
      \tyatomic{x_2}{t_2} \vdash \cr
      \egive{x_1}{x_2}: \tybar{}{&
        \tyatomic{x_2}{t_2}
      }
      \end{array}
    }

    \inferrule[Take]{
    t_2 \text{ adopts } t_1
    }{
      \begin{array}{r@{\;}l}
      K;
      \tyatomic{x_1}{\tydynamic}\ \ast&
      \tyatomic{x_2}{t_2} \vdash \cr
      \etake{x_1}{x_2}: \tybar{}{
        \tystar
        {\tyatomic{x_1}{t_1}}
        {&\tyatomic{x_2}{t_2}}
      }
      \end{array}
    }

    \inferrule[Fail]
    {}{
      K; P \vdash \efail : t
    }

    \inferrule[Sub]
    {K; P_2 \vdash e: t_1 \\\\ P_1 \leq P_2 \\ t_1 \leq t_2}
    {K; P_1 \vdash e: t_2}

    \inferrule[Frame]
    {K; P_1 \vdash e: t}
    {K; P_1\ast P_2 \vdash e: \tybar{t}{P_2}}

    \inferrule[ExistsElim]{
      K, X: \kappa; P \vdash e : t
    }{
      K; \tyexists X \kappa P \vdash e : t
    }
  \end{mathpar}
  \caption{Typing rules}
  \label{fig:typing-rules}
\end{figure*}

\begin{figure*}
\begin{mathpar}
    \inferrule[LetTuple]
    {(\vec{t}\,) \text{ is duplicable} \\\\
     K, \vec{x}: \kterm; P \ast \tyatomic{x}{(\vec{t}\,)}
                           \ast \tyatomic{\vec{x}}{\vec{t}}
     \vdash e: t}
    {K; P\ast \tyatomic{x}{(\vec{t}\,)} \vdash \elet{(\vec{x})}{x}{e}: t}

    \inferrule[LetDataMatch]
    {(\vec{t}\,) \text{ is duplicable} \\\\
     K, \vec{x}: \kterm;
     P \ast \tyatomic{x}{\tyconcreteadopts{A}{\vec{f}:\vec{t}}{u}}
       \ast \tyatomic{\vec{x}}{\vec{t}}
     \vdash e: t}
    {K; P\ast \tyatomic{x}{\tyconcreteadopts{A}{\vec{f}:\vec{t}}{u}}
     \vdash
     \elet{\econstruct{A}{\vec{f}=\vec{x}}}{x}{e}: t}

    \inferrule[LetDataMismatch]
    {\text{\kw{A} and \kw{B} belong to a common algebraic data type}}
    {K; P\ast \tyatomic{x}{\tyconcreteadopts{A}{\vec{f}:\vec{t}}{u}}
     \vdash
     \elet{\econstruct{B}{\vec{f'}=\vec{x}}}{x}{e}: t}

    \inferrule[LetDataUnfold]
    {\kern1.2mm
     \tyatomic{x}{\tyconcreteadopts{A}{\vec{f}:\vec{t}}{u}}
     \text{ is an unfolding of }
     \tyapp{T}{\vec{T}} \\\\
     K;
     P \ast \tyatomic{x}{\tyconcreteadopts{A}{\vec{f}:\vec{t}}{u}}
     \vdash
     \elet{\econstruct{A}{\vec{f}=\vec{x}}}{x}{e}: t}
    {K; P\ast \tyatomic{x}{\tyapp{T}{\vec{T}}}
     \vdash
     \elet{\econstruct{A}{\vec{f}=\vec{x}}}{x}{e}: t}    
\end{mathpar}
\caption{Auxiliary typing rules for pattern matching}
\label{fig:pat}
\end{figure*}

\begin{figure*}
  \centering
  \begin{mathpar}
    \begin{longue}
    \inferrule[Reflexive]
    {}
    {P \leq P}

    \inferrule[Transitive]
    {P_1 \leq P_2 \\ P_2 \leq P_3}
    {P_1 \leq P_3}

    \inferrule[EmptyTop]
    {}
    {P \leq \tyempty}

    \inferrule[EmptyAppears]
    {}
    {P \leq \tyempty \ast P}

    \inferrule[StarCommutative]
    {}
    {P_1 \ast P_2 \leq P_2 \ast P_1}    

    \inferrule[StarAssociative]
    {}
    {P_1 \ast (P_2 \ast P_3) \leq (P_1 \ast P_2) \ast P_3}

    \end{longue}
    \inferrule[EqualityReflexive]
    {} 
    {\tyempty \leq (x = x)}

    \inferrule[EqualsForEquals]
    {}
    {
      \tystar
      {(y_1 = y_2)}
      {[y_1/x]P}
      \subsub
      \tystar
      {(y_1 = y_2)}
      {[y_2/x]P}
    }
  
    \inferrule[Duplicate]
    {P\text{ is duplicable}}
    {P \leq P * P}
    \begin{longue}

    \inferrule[HideDuplicablePrecondition]
    {P \text{ is duplicable}}
    {
      (\tyatomic{x}{\tyarrow{\tybar{t_1}{P}}{t_2}})
      \ast P
      \leq
      \tyatomic{x}{\tyarrow{t_1}{t_2}}
    }
    \end{longue}

    \inferrule[MixStar]
    {}
    {\tyatomic{x}{t} \ast P \subsub \tyatomic{x}{\tybar{t}{P}}}

    \inferrule[Weaken] 
    {}
    {P_1 \ast P_2 \leq P_2}

    \inferrule[ExistsIntro]
    {}
    {[T/X]P \leq \tyexists X\kappa P}
    \begin{longue}

    \inferrule[ExistsStar]
    {}
    {P_1 \ast \tyexists X\kappa{P_2} \subsub \tyexists X\kappa{(P_1 \ast P_2)} }
    \end{longue}

    \inferrule[ExistsAtomic]
    {}
    {
      \tyatomic{x}{\tyexists X\kappa t}\\\\
      \subsub
      \tyexists X\kappa{(\tyatomic{x}{t})}
    }

    \begin{longue}
    \inferrule[DecomposeTuple]
    {}
    {
    \begin{array}{r@{}l}
    & \tyatomic{y}{(\ldots,t,\ldots)} \cr
    \subsub
    \tyexists x\kterm{(
      \tystar
        {& \tyatomic{y}{(\ldots, \tysingleton{x}, \ldots)}}
        {\tyatomic{x}{t}}
      )}
    \end{array}
    }

    \end{longue}
    \inferrule[DecomposeBlock]
    {}
    {
      \begin{array}{r@{}l}
        & \tyatomic{y}{\tyconcreteadopts{A}{F[f:t]}{u}} \cr
        \subsub
        \tyexists x\kterm{(
          \tystar
          {& \tyatomic{y}{\tyconcreteadopts{A}{F[f=x]}{u}}}
          {\tyatomic{x}{t}}
        )}
      \end{array}
    }

    \inferrule[Fold]
    {\tyconcreteadopts{A}{\vec{f}: \vec{t}}{u} \text{ is an unfolding of } \tyapp{T}{\vec{T}}}
    {
      \tyatomic{x}{\tyconcreteadopts{A}{\vec{f}: \vec{t}}{u}}
      \leq
      \tyatomic{x}{\tyapp{T}{\vec{T}}}
    }

    \inferrule[Unfold]
    {\tyconcreteadopts{A}{\vec{f}: \vec{t}}{u} \text{ is an unfolding of } \tyapp{T}{\vec{T}} \\\\
     \tyapp{T}{\vec{T}} \text{ has only one branch}}
    {
      \tyatomic{x}{\tyapp{T}{\vec{T}}}
      \leq
      \tyatomic{x}{\tyconcreteadopts{A}{\vec{f}: \vec{t}}{u}}
    }

     \inferrule[DynamicAppears]
     {t\text{ is exclusive}}
     {\tyatomic{x}{t} \leq \tyatomic{x}{t} \ast
     \tyatomic{x}{\tydynamic}}
   \begin{longue}

   \inferrule[CoArrow]
   {u_1 \leq t_1 \\ t_2 \leq u_2}
   {\tyatomic{x}{\tyarrow{t_1}{t_2}} \leq \tyatomic{x}{\tyarrow{u_1}{u_2}}}

   \inferrule[CoTuple]
   {\vec{t} \leq \vec{u}}
   {\tyatomic{x}{(\vec{t}\,)} \leq \tyatomic{x}{(\vec{u})}}

   \inferrule[CoBlock]
   {\vec{t} \leq \vec{u} \\ t \leq u}
   {\tyatomic{x}{\tyconcreteadopts{A}{\vec{f}:\vec{t}}{t}} \leq \tyatomic{x}{\tyconcreteadopts{A}{\vec{f}:\vec{u}}{u}}}

   \inferrule[CoStar]
   {P_1 \leq P_2 \\ Q_1 \leq Q_2}
   {P_1 \ast Q_1 \leq P_2 \ast Q_2}
   \end{longue}
  \end{mathpar}
   \makeatletter
   \caption{Permission subsumption \@ifundefined{fplong}{(not all rules shown)}{}}
   \makeatother
  \label{fig:sub}
\end{figure*}

\section{Type assignment}
\label{sec:typing}

\subsection{The typing judgment}

The typing judgment takes the form $K; P \vdash e: t$. It is inductively
defined in Figures~\ref{fig:typing-rules} and~\ref{fig:pat}. The kind
environment $K$ maps variables to kinds. This judgment means that, by
consuming the permission~$P$, the expression $e$ produces a value of type
$t$. It is analogous to a Hoare logic or separation logic triple, where $P$ is
the precondition and $t$ is the postcondition.

The typing judgement relies on a well-kindedness judgement of the form $K
\vdash t: \kappa$. It ensures that types are
well-kinded and that the syntactic facilities of the surface syntax
(\sref{sec:surface-syntax}) are used properly. For conciseness, in the typing
rules, we omit all freshness and well-kindedness side conditions.

The typing rules require many sub-expressions to be variables. For instance,
the rule \Rule{Read} cannot handle a field access expression of the form
$e.f$: instead, it requires $x.f$. This requirement is met by first performing
a monadic transformation, which introduces extra \kw{let} constructs.
Furthermore, the pattern matching rules (\fref{fig:pat}) cannot handle deep
patterns: they require shallow patterns. Again, this requirement is met by
introducing extra \kw{let} constructs. We omit the details of these
transformations.

\Rule{Var} is the axiom rule. It is worth noting that, in conjunction with
the subsumption rule \Rule{EqualityReflexive} (\fref{fig:sub}), it
allows proving that $x$ has type $\tysingleton{x}$, even in the absence of
any hypothesis about $x$.

\Rule{Let} corresponds to the sequence rule of separation logic.

\Rule{Function} states that a \emph{duplicable} permission $P$ that exists at
the function definition site is also available within the function body.
Requiring $P$ to be duplicable allows us to consider every function type
duplicable. Thus, a function can be shared without restriction and can be
invoked as many times as desired, provided of course that one is able to
satisfy its precondition. If one wishes to write a function that captures
a non-duplicable permission $P$, and can be invoked at most once, this is
still possible. Indeed, a type $t_1\osf t_2$ of
``one-shot'' functions can be defined as:
$$
\tyexists{p}{\kperm}{\tybar{(\tyarrow{\tybar{t_1}{p}}{t_2})}{p}}
$$
This is a conjunction of a function whose precondition is $p$ and of one copy
of $p$. Because $p$ is abstract, it is considered affine. Hence, at most one
call is possible, after which $p$ is consumed and the function becomes
unusable.

\Rule{Application} corresponds to the rule for procedure calls in separation
logic. The caller gives up the permission $\tyatomic{x_2}{t_2}$, which is
consumed, and in return gains a permission for the result of the
function call, at type $t_1$. In other words, because types have an ownership
reading, a function type $t_1\rightarrow t_2$ describes not only the shape of
the function's arguments and results, but also the side effects that the
function may perform, as well as the transfers of ownership that occur from
the caller to the callee and back.

\Rule{New} uses a structural type to describe the newly-allocated memory
block in an exact manner. \Rule{Tuple} is analogous.

\Rule{Read} requires a structural permission $\tyatomic{x}{\tyconcrete{A}{F[f:
    t]}}$, which guarantees that $x$ points to a memory block that contains a
field named $f$, and allows us to dereference $x.f$\footnote{We write $F[f:t]$
  for a sequence of field/type pairs in which the pair $f:t$ occurs. The
  \kw{adopts} clause, if present, is irrelevant. We overload field
names: there could exist multiple data constructors with a field named~$f$.
There can be at most one permission of the form
$\tyatomic{x}{\tyconcrete{A}{F[f: t]}}$, though, which allows
disambiguation to take place.}.
\Rule{Read} concludes that the field access expression $x.f$ has type $t$, and
that the structural permission $\tyatomic{x}{\tyconcrete{A}{F[f: t]}}$ is
preserved. There is a catch: because the type~$t$ occurs twice in this
postcondition, we must require $t$ to be duplicable, or the rule would
be unsound. Fortunately, this is not a problem: by using \Rule{DecomposeBlock}
(\fref{fig:sub}; also explained earlier, see \sref{sec:example:permissions}
and \sref{sec:extent}), it is possible to arrange for $t$ to be a singleton
type, which is duplicable.

Like \Rule{Read}, \Rule{Write} requires a structural permission, of the form
$\tyatomic{x_1}{\tyconcrete{A}{F[f: t_1]}}$. It checks that this permission is
exclusive, i.e., the data constructor~$\kw{A}$ is associated with a
mutable algebraic data type. This ensures that we have write access. In fact,
since we have exclusive access to $x_1$, a strong (type-changing) update is
sound. The permission is changed to
$\tyatomic{x_1}{\tyconcrete{A}{F[f: t_2]}}$, where $t_2$ is the type of $x_2$.
Without loss of generality, one may let $t_2$ be the singleton type
$\tysingleton{x_2}$. This allows the type-checker to record that $x_1.f$ and
$x_2$ are now aliases. If desired, the permissions
$\tyatomic{x_1}{\tyconcrete{A}{F[f = x_2]}}$ and $\tyatomic{x_2}{t_2}$ can
later be combined by \Rule{DecomposeBlock} to yield
$\tyatomic{x_1}{\tyconcrete{A}{F[f: t_2]}}$. Because \Rule{DecomposeBlock},
read from right to left,
involves a loss of information, it is typically applied by the type-checker
only ``on demand'', i.e., to satisfy a function postcondition or a type
annotation.

\Rule{Match} is used to type-check a case analysis construct. Each branch is
type-checked independently. We currently do not check that the case analysis
is exhaustive, but plan to add this feature in the future.
The premise relies on a judgment of the form $K; P \vdash \elet pxe : t$.
This is not a new judgment; it is an ordinary typing judgement, but, for
clarity, the typing rules that have a conclusion of this form are
isolated in \fref{fig:pat}. Although these rules may appear somewhat daunting,
they are in fact quite straightforward. \Rule{LetTuple} checks that $x$ is a
tuple, i.e., we have a permission of the form
$\tyatomic{x}{\tytuple{t_1,\ldots,t_n}}$.  If that is the case, then
matching~$x$ against the tuple pattern $\etuple{x_1,\ldots,x_n}$ is permitted, and
gives rise to a conjunction of permissions of the form
$\tyatomic{x_i}{t_i}$. Because the permission for $x$ is not lost, the
types~$t_i$ are duplicated, so they are required to be duplicable. Again, this
requirement causes no loss of generality, since one can arrange to introduce
singleton types ahead of time.
\Rule{LetDataMatch} is analogous to \Rule{LetTuple}, but concerns a (mutable
or immutable) memory block. \Rule{LetDataMismatch} concerns the situation
where the pattern, which mentions the data constructor \kw{B}, will clearly
not match $x$, which is statically known to have the tag \kw{A}. In that case,
the branch is dead code, and is considered well-typed. \Rule{LetDataUnfold}
\emph{refines} a nominal permission, such as $\tyatomic{x}{\tylist{a}}$, by
replacing it with a structural one, such as
$\tyatomic{x}{\tyconcrete{Cons}{\fhead:a;\ftail:\tylist{a}}}$,
obtained by unfolding the algebraic data type and specializing it with
respect to the data constructor that appears in the pattern. We omit
the exact definition of unfolding.

\Rule{WriteTag} type-checks a \emph{tag update} instruction, which modifies
the tag carried by a memory block. Like \Rule{Write}, it requires an exclusive
permission for this block. It further requires the new tag~\kw{B} to carry the
same number of fields as the previous tag~\kw{A}. (Thus, the block does not
have to be enlarged or shrunk.) The structural permission is updated in a
straightforward way. The types~$\vec{t}$ of the fields do not change. The
names of the fields change from $\vec{f}$ to $\vec{f'}$, where the sequences
of fields are ordered in the same way as in the (user-provided) definitions of
$\kw{A}$ and $\kw{B}$. It is worth noting that $\kw{A}$ and~$\kw{B}$ need not
belong to the same algebraic data type: thus, a memory block can be re-used
for a completely new purpose. Furthermore, the tag $\kw{B}$ may be associated
with an immutable algebraic data type: in that case, the block is
\emph{frozen}, that is, becomes forever immutable. This feature is exploited
in the concatenation of immutable lists (\fref{fig:list},
\lineref{line:tagof}).

\Rule{Give} and \Rule{Take} are explained later on (\sref{sec:adoption}).

\Rule{Sub} is analogous to Hoare's rule of consequence. It relies on
permission subsumption, $P_1\leq P_2$, defined in \fref{fig:sub} and
discussed further on (\sref{sec:sub}),
and on subtyping, $t_1\leq t_2$, defined as
$\tyatomic{x}{t_1}\leq\tyatomic{x}{t_2}$ for a fresh~$x$.

\Rule{Frame} is analogous to the frame rule of separation logic.

\subsection{The permission subsumption judgment}
\label{sec:sub}

\begin{courte}
A subset\footnote{The reader is referred to the extended version of this
paper~\cite{self-long} for the full set of rules. The omitted rules include:
conjunction is commutative, associative, and has unit $\tyempty$; covariance
and contravariance of the type constructors; and a few more.}
of the subsumption rules appears in \fref{fig:sub}.
\end{courte}
\begin{longue}
The rules that define the subsumption judgment appear in \fref{fig:sub}.
We comment a subset of them. 
\end{longue}
Since $x = y$ is sugar for $\tyatomic{x}{\tysingleton{y}}$, the rule
\Rule{EqualityReflexive} can be understood as a claim that $x$ inhabits
the singleton type $\tysingleton{x}$. \Rule{EqualsForEquals} shows how
equations are exploited: if $y_1$ and~$y_2$ are known to be equal, then they
are interchangeable. (We write $\subsub$ for subsumption in both
directions.) \Rule{Duplicate} states that a permission that is syntactically
considered duplicable can in fact be duplicated. \Rule{MixStar} introduces and
eliminates $\tybar{t}{P}$. \Rule{Weaken} states that every permission is affine.
\Rule{ExistsIntro} introduces an existential permission; \Rule{ExistsAtomic}
converts between an existential permission and
an existential type. When read from left to right, \Rule{DecomposeBlock},
which was discussed earlier
(\sref{sec:example:permissions}, \sref{sec:extent}), introduces a fresh
name~$x$ for the value stored in $y.f$. When read from right to left, it
forgets such a name. (In that case, it is typically used in conjunction
with \Rule{ExistsIntro}.) This rule is related to Sing\#'s explicit
``expose''~\cite{faehndrich-singularity-06}.
\Rule{Fold} folds an algebraic data type definition,
turning a structural type into a nominal type. Unfolding is normally
performed by case analysis (see \Rule{LetDataUnfold}
in \fref{fig:pat}), but in the special case where an algebraic data
type has only one branch (i.e., it is a record type), it can be implicitly
unfolded by \Rule{Unfold}. \Rule{DynamicAppears} is explained later on
(\sref{sec:adoption}).

\subsection{The \mezzo type-checker}
\label{sec:type-checker}

Implementing a checker for a highly non-deterministic system such as \mezzo
poses technical challenges. Our type-checker greedily introduces fresh
auxiliary names so as to ``normalize'' the types and permissions at hand. A
persistent union-find data structure keeps track of the permissions of the
form ``$x = y$''. The use of flexible variables enables a limited form of type
inference. At present, the implemented type-checker is not complete with
respect to the type assignment rules. This is only a brief overview: a
follow-up paper devoted to the implementation of \mezzo is in the works.

\section{Surface syntax}
\label{sec:surface-syntax}

The internal syntax, which we have been using so far, can be fairly
verbose. To remedy this, we introduce two syntactic conventions, which rely on
the \emph{name introduction} construct and on the \kw{consumes} keyword
(\fref{fig:mezzo-types}). Two transformations
eliminate these constructs, so as to obtain a type expressed in the internal
syntax.
\begin{courte}
They are formalized in the extended version of the present
paper~\cite{self-long}. Here, we give only an intuition.
\end{courte}
\begin{longue}
This section contains an informal discussion that gives the intuition for these
transformations. The two transformations are formalized and discussed in
\ref{sec:syntax-appendix}.
\end{longue}

  \subsection{The name introduction form}

  The construct $\tynameintro{x}{t}$ allows introducing a name
  $x$ for a component of type $t$. This allows writing ``dependent function
  types'', such as $\tyarrow{(\tynameintro{x_1}{t_1})}{(\tynameintro{x_2}{t_2})}$, where
  by convention $x_1$ is bound within $t_1$ and $t_2$, while $x_2$ is bound
  within~$t_2$. This is desugared by quantifiying $x_1$ universally \emph{above}
  the arrow and quantifying $x_2$ existentially in the right-hand side of the arrow.

  As an example, consider the type of \li|:=|, the function that writes a
  reference. This function expects a pair of a
  reference~$x$ whose content has type~$a$ and of a value of
  type $b$, which it stores into $x$. At the end, $x$ has become a reference
  whose content has type~$b$. The variable $x$ must be mentioned in the pre-
  and postcondition. In the internal syntax, the type of \li|:=| is:
  \[
  \forall a,b.
  \tyforall{x}{\kterm}
  {\tyarrow
    {({\typackage{x}{\kw{ref}\ a}}, {b})}
    {\\\tybar{}{\tyatomic{x}{\kw{ref}\ b}}}}
  \]
  Thanks to the name introduction form, instead of planning ahead
  and quantifying $x$ in front of the function type, one
  names the first argument ``$x$'' on the fly. Thus, in the surface syntax,
  one writes:
  $$
  \forall a,b.
  \tyarrow
  {(\tyconsumes{\tynameintro{x}{\kw{ref}\ a}}, \tyconsumes{b})}
  {\tybar{}{\tyatomic{x}{\kw{ref}\ b}}}
  $$
  This is not significantly shorter, because of the $\kw{consumes}$
  keyword, which must be used in the surface syntax, as explained
  below. In actual use, though, the comfort afforded by this feature
  is critical.

  \subsection{The consumes annotation}

  Often, a permission is required \emph{and returned} by a function, in which
  case it is unpleasant to have to write this permission twice, in the
  precondition and postcondition. Drawing inspiration from
  Sing\#~\cite{faehndrich-singularity-06}, we adopt the convention that, in
  the surface syntax, \emph{by default, the permission for the argument is
    required and returned}, i.e., it is \emph{not} consumed.

For instance, the type of the list \li|length|
function, which in the internal syntax is:
$$
\forall a.
\forall (x: \kterm)
\tyarrow
{\tybar{\tysingleton{x}}{\tyatomic{x}{\tylist{a}}}}
{\tybar{\tyint}{\tyatomic{x}{\tylist{a}}}}
$$
can in the surface syntax be written in a much more pleasant form:
$$\forall a.
  \tyarrow
    {\tylist{a}}
    {\tyint}
$$
The type $\tylist{a}$ is mentioned once, instead of twice, and as a side
effect, the need to name the argument~$x$ vanishes.
  
When a permission \emph{is} consumed, though, we need a way of indicating
this. This is the purpose of the \li|consumes| keyword. When a component is
marked with this keyword, the permission for this component is required and
not returned.  This keyword makes sense only in the left-hand side of an
arrow.

  Since internal syntax and surface syntax interpret the function type differently,
  a translation is required, regardless
  of whether \li|consumes| is used. Consider a function type of the form
  $\tyarrow{t}{u}$, where
  $t$ does not contain any name introduction forms.
  Let $t_1$ stand for $[\tau/\tyconsumes{\tau}]t$, i.e.,
  a copy of $t$ where the \kw{consumes} keyword is erased.
  Let $t_2$ stand for
  $[\top/\tyconsumes\tau]t$, i.e.,
  a copy of $t$ where every component marked with this keyword is replaced
  with $\top$%
  \footnote{
    Here, we write $\top$ for $\tyunknown$ or $\tyempty$,
    depending on whether the \li|consumes| keyword is applied to a type or a
    permission.
  }.
  Then, the translation of 
  this function type
  is
  $\tyarrow{(\tynameintro{x}{t_1})}{\tybar{u}{\tyatomic{x}{t_2}}}$.
  The parts of the argument that are \emph{not} marked as
  consumed are returned to the caller.

  The type of the function \li|insert|, which appears in
  \fref{fig:intf} and is discussed in
  \sref{sec:adoption-overview}, states that the first argument
  is consumed, while the second argument is not. Its translation
  into the internal syntax is as follows:
  \[\begin{array}{l}
  \tyforall{a}{\ktype}{
    \tyforall{x}{\kterm}{\\
      \tyarrow{
        \typackage{x}{
          (a, \kw{bag } a)
        }
      }{
        \tybar{}{\tyatomic{x}{(\tyunknown, \kw{bag } a)}}
      }
    }
  }
  \end{array}\]

  \subsection{Function definitions}

  In the internal syntax, functions take the form $\efun{(x: t_1)}{t_2}{e}$, where one
  variable, namely $x$, is bound in $e$.
  In the surface syntax, instead, functions take the form
  $\efun{t_1}{t_2}{e}$.
  The argument type~$t_1$ is interpreted as a pattern, and the names that it
  introduces 
  are considered bound in $e$. An example is
  $\efun{(x:\tyint,y:\tyint)}{\tyint}{x+y}$, where $(x: \tyint, y: \tyint)$ is the type of the
  argument, $\tyint$ is the type of the result, and $x$ and $y$ are bound in the
  function body, which is $x+y$.

\section{Adoption and abandon}
\label{sec:adoption}

\begin{figure}[t]
\begin{lstlisting}[numbers=left]
abstract bag a
val create: [a] () -> bag a
val insert: [a] (consumes a, bag a) -> ()
val retrieve: [a] bag a -> option a
\end{lstlisting}
\caption{An interface for bags}
\label{fig:intf}
\end{figure}

\begin{figure}[t]
\begin{lstlisting}[numbers=left]
data mutable cell a =
  Cell { elem: a; next: dynamic } -- \label{line:next:dynamic}

data mutable bag a =
     Empty { head, tail: () }
| NonEmpty { head, tail: dynamic } -- \label{line:head-tail:dynamic}
adopts cell a -- \label{line:adopts}

val create [a] () : bag a =
  Empty { head = (); tail = () }

val insert [a] (consumes x: a, b: bag a) : () =
  let c = Cell { elem = x; next = () } in
  c.next <- c;
  give c to b; -- \label{line:adopt}
  match b with -- \label{line:match:one}
  | Empty ->
      tag of b <- NonEmpty; -- \label{line:empty-to-nonempty:begin}
      b.head <- c; -- \label{fig:write:c:1}
      b.tail <- c -- \label{line:empty-to-nonempty:end} \label{fig:write:c:2}
  | NonEmpty ->
      take b.tail from b; -- \label{line:focus:entry}
      b.tail.next <- c; -- \label{fig:write:c:3}
      give b.tail to b; -- \label{line:focus:exit}
      b.tail <- c -- \label{fig:write:c:4}
  end

val retrieve [a] (b: bag a) : option a =
  match b with -- \label{line:match:two}
  | Empty ->
      None
  | NonEmpty ->
      take b.head from b; -- \label{line:abandon}
      let x = b.head.elem in -- \label{line:head:elem}
      if b.head == b.tail then begin
        tag of b <- Empty; -- \label{line:nonempty-to-empty:begin}
        b.head <- ();
        b.tail <- () -- \label{line:nonempty-to-empty:end}
      end else begin
        b.head <- b.head.next -- \label{line:head:next}
      end;
      Some { value = x }
  end
\end{lstlisting}
\caption{A FIFO implementation of bags}
\label{fig:fifo}
\end{figure}

The permission discipline that we have presented so far has limited expressive
power. It can describe immutable data structures with arbitrary sharing and
tree-shaped mutable data structures. However, because mutable memory blocks
are controlled by exclusive permissions, it cannot describe mutable data
structures with sharing.

\subsection{Overview}
\label{sec:adoption-overview}

In order to illustrate this problem, let us imagine how one could implement a
``bag'' abstraction. A bag is a mutable container, which supports two
operations: inserting a new element and retrieving an arbitrary element.

We would like our implementation to offer the interface in \fref{fig:intf}.
There, \li|bag| is presented as an abstract type. Because it is not explicitly
declared duplicable, it is regarded as affine. Hence, a bag ``has a unique
owner'', i.e., is governed by a non-duplicable permission. The
function \li|create| creates a new bag, whose ownership is transferred to the
caller. The type of \li|insert| indicates that
\li|insert(x, b)| requires the permissions \qli|x @ t| and \qli|b @
bag t|, for some type~\li|t|, and returns only the latter. Thus, the caller
gives up the ownership of \li|x|, which is ``transferred to the
bag''. Conversely, the call \qli|let o = retrieve b in ...| produces the
permission \qli|o @ option a|, which means that the ownership of the retrieved
element (if there is one) is ``transferred from the bag to the caller''.

To implement bags, we choose a simple data structure, namely a mutable
singly-linked list. One inserts elements at the tail and extracts elements at
the head, so this is a FIFO implementation. One distinguished object~\li|b|,
``the bag'', has pointers to the head and tail of the list, so as to allow
constant-time insertion and extraction. (We use ``object'' as a synonym for
``memory block''.)

This data structure is not tree-shaped: the last cell in the list is
accessible via two distinct paths. In order to type-check this code, we must
allow the ownership hierarchy and the structure of the heap to differ.  More
specifically, we would like to view the list cells as collectively owned by
the bag~\li|b|. That is, we wish to keep track of just one exclusive
permission for the \emph{group} formed by the list cells, as opposed to one
permission per cell.

We use the name~\li|b| as a name for this group. When a cell~\li|c| joins the
group, we say that \li|b| \emph{adopts}~\li|c|, and when
\li|c| leaves the group, we say that \li|b| \emph{abandons} \li|c|. In other
words, the bag~\li|b| is an \emph{adopter}, and the list cells~\li|c| are
its \emph{adoptees}. In terms of ownership, adopter and adoptees form a unit:
the exclusive permission that controls~\li|b| also represents the ownership of
the group, and is required by the adoption and abandon operations.

Adoption requires and consumes an exclusive permission for the cell~\li|c|
that is about to be adopted: the ownership of~\li|c| is transferred to the
group. Conversely, abandon produces an exclusive permission for the
cell~\li|c| that is abandoned: the group relinquishes the ownership of~\li|c|.

Abandon must be carefully controlled. If a cell could be abandoned twice, two
permissions for it would appear, which would be unsound. Due to aliasing,
though, it is difficult to statically prevent this problem. Instead, we decide
to record \emph{at runtime} which object is a member of which group, and to
verify \emph{at runtime} that abandon is used in a safe way.

\subsection{Details}
\label{sec:adoption:details}

Let us now explain in detail the dynamic semantics of adoption and abandon
(what these operations do) as well as their static semantics (what the
type-checker requires).

\paragraph{Adopter fields}
\label{sec:adoption:details:adopter}

We maintain a pointer from every adoptee to its adopter. Within every object,
there is a hidden \qli|adopter| field, which contains a pointer to the
object's current adopter, if it has one, and \li|null| otherwise. This
information is updated when an object is adopted or abandoned.
In terms of space, the cost of this design decision is one field per
object.

\paragraph{The type dynamic}
\label{sec:adoption:details:dynamic}

The permission \qli|c @ dynamic| guarantees that \li|c| is a pointer to a
memory block
\begin{longue}
(as opposed to, say, an integer value, or a function value)
\end{longue}
and grants read access to the field \li|c.adopter|. This can be used to verify
the identity of \li|c|'s adopter. In other words, \qli|c @ dynamic| can be
viewed as a permission to perform a \emph{dynamic group membership test}. It
is a duplicable permission. It appears spontaneously when \li|c| is known to
be a (mutable) object: this is stated by the rule \Rule{DynamicAppears}
in \fref{fig:sub}.

In the bag implementation, shown in \fref{fig:fifo}, the \li|head| and
\li|tail| fields of a non-empty \li|bag| object, as well as the \li|next|
field of every \li|cell| object, have type \li|dynamic|
(lines~\ref{line:next:dynamic} and~\ref{line:head-tail:dynamic}).  Because
\li|dynamic| is duplicable, sharing is permitted: for instance, the
pointers \li|b.head| and \li|b.tail| might happen to be equal.

\paragraph{Adopts clauses}
\label{sec:adoption:details:adopts}

When a cell~\li|c| is adopted, the exclusive permission that describes it,
namely \qli|c @ cell a|, disappears. Only \qli|c @ dynamic| remains. As a
result, the information that \li|c| is a cell is lost: the type-checker can no
longer tell how many fields exist in the object~\li|c| and what they
contain. When the bag~\li|b| later abandons~\li|c|, we would like the
permission \qli|c @ cell a| to re-appear. How can the type-checker recover
this information?

Fortunately, when \li|b| abandons \li|c|, the type-checker has access to the
type of~\li|b|. Thus, provided the type of the adopter determines the type of
its adoptees, this problem is solved.

For an object \li|b| of type \li|t| to serve as an adopter, where \li|t| is an
algebraic data type, we require that the definition of \li|t| contain the
clause \qli|adopts u| and that \li|t| and \li|u| be exclusive types. This is
illustrated in \fref{fig:fifo}, where the definition of \qli|bag a|
says \qli|adopts cell a| (line~\ref{line:adopts}).

Because the type of the adoptees must not be forgotten when an algebraic data
type is unfolded, structural permissions also carry an \li|adopts| clause.
In the case of bags, for instance, the permission \qli|b @ bag a| is refined by
the \li|match| constructs of lines~\ref{line:match:one}
and~\ref{line:match:two} into either
``\li|b @ Empty { head, tail: () } adopts cell a|'' or
``\li|b @ NonEmpty { head, tail: dynamic } adopts cell a|'',
and, conversely, either of these permissions can be folded back to \qli|b @ bag a|.

We write that ``\li|t| adopts \li|u|'' if either \li|t| is an algebraic data
type whose definition contains the clause \qli|adopts u| or
\li|t| is a structural type that contains the clause \qli|adopts u|.

\paragraph{Adoption}
\label{sec:adoption:details:adoption}

The syntax of adoption is \qli|give c to b|. 
This instruction requires two permissions \qli|c @ u| and \qli|b @ t|,
where \li|t| adopts~\li|u| (\Rule{Give}, \fref{fig:typing-rules}).
At the program point that follows this instruction, the permission \qli|b @ t|
remains available, but \qli|c @ u| has been consumed. Fortunately, not
everything about \li|c| is forgotten. The permission \qli|c @ dynamic|, which
is present before the adoption instruction because \qli|c @ u| spontaneously
gives rise to \qli|c @ dynamic|, remains present after adoption.

The runtime effect of this operation is to write the address~\li|b| to the
field \li|c.adopter|. The exclusive permission \qli|c @ u|
guarantees that this field exists and that its value, prior to adoption, is
\li|null|.

In the bag implementation (\fref{fig:fifo}), adoption is used at the
beginning of \li|insert| (line~\ref{line:adopt}), after a fresh cell~\li|c|
has been allocated and initialized. This allows us to maintain the (unstated)
invariant that every cell that is reachable from \li|b| is adopted by~\li|b|.

\paragraph{Abandon}
\label{sec:adoption:details:abandon}

The syntax of abandon is \qli|take c from b|.
This instruction requires \qli|b @ t| and \qli|c @ dynamic|,
where \li|t| adopts~\qli|u|, for some type~\li|u| (\Rule{Take}, \fref{fig:typing-rules}).
After this instruction, \qli|b @ t|
remains available. Furthermore, the permission \qli|c @ u| appears.

The runtime effect of this operation is to check that the field \li|c.adopter|
contains the address~\li|b| and to write \li|null| into this field, so as to
reflect the fact that \li|b| {abandons}~\li|c|. If this check fails, the
execution of the program is aborted.

In the bag implementation (\fref{fig:fifo}), abandon is used near the
beginning of \li|retrieve|, at line~\ref{line:abandon}. There, the first cell
in the queue, \li|b.head|, is abandoned by \li|b|. This yields a
permission at type \qli|cell a| for this cell. This permission lets us read
\li|b.head.elem| and \li|b.head.next| and allows us to produce the permission
\qli|x @ a|, where \li|x| is the value found in \li|b.head.elem|.

Abandon and adoption are also used inside \li|insert|, at
lines~\ref{line:focus:entry} and~\ref{line:focus:exit}. There, the bag \li|b|
is non-empty, and the cell \li|b.tail| must be updated in order to reflect the
fact that it is no longer the last cell in the queue. However, we cannot just
go ahead and access this cell, because the only permission that we have at
this point for this cell is at type \qli|dynamic|. Instead, we must take the
cell out of the group, update it, and put it back. We allow writing
\qli|taking b.tail from b begin ... end| as sugar for such a
well-parenthesized use of \li|take| and \li|give|.

\subsection{Discussion}

To the best of our knowledge, adoption and abandon are new.  Naturally, the
concept of group, or region, has received sustained interest in the
literature~\cite{crary-cc-99,deline-faehndrich-01,faehndrich-deline-02,swamy-06}.
Regions are usually viewed either as a dynamic memory management mechanism or
as a purely static concept. Adoption and abandon, on the other hand, offer a
dynamic ownership control mechanism, which complements our static permission
discipline.

Adoption and abandon are a very flexible mechanism, but also a dangerous one.
Because abandon involves a dynamic check, it can cause the program to
encounter a fatal failure at runtime. In principle, if the programmer knows
what she is doing, this should never occur. There is some danger, but that is
the price to pay for a simpler static discipline. After all, the danger is
effectively less than in ML or Java, where a programming error that creates an
undesired alias goes completely undetected---until the program misbehaves in
one way or another.

One might wonder why the type $\tydynamic$ is so uninformative: it gives no
clue as to the type of the adoptee or the identity of the adopter. Would it be
possible to parameterize it so as to carry either information? The short
answer is negative. The type $\tydynamic$ is duplicable, so the information
that it conveys should be stable (i.e., forever valid). However, the type of
the adoptee, or the identity of the adopter, may change with time, through a
combination of strong updates and \li|give| and \li|take| instructions. Thus,
it would not make sense for $\tydynamic$ to carry more information.

That said, we believe that adoption and abandon will often be used according
to certain restricted protocols, for which more information is stable, hence
can be reflected at the type level. For instance, in the bag implementation, a
cell only ever has one adopter, namely a specific bag~$b$. In that case, one
could hope to work with a parameterized type $\tydynamicprime{b}$, whose
meaning would be ``either this object is currently not adopted, or it is
adopted by~$b$''. Ideally, $\kw{dynamic}'$ would be defined on top of
$\tydynamic$ in a library module, and its use would lessen the risk of
confusion.

\begin{longue}

\paragraph{Tag update}
\label{sec:tag:update:comments}

Our implementation of bags exploits the fact that it is permitted to mutate
not just the fields, but also the tag of a mutable object. An object of
type \qli|bag a| carries either the tag \li|Empty|, in which case the
\li|head| and \li|tail| fields have the unit type~\li|()|, or the tag
\li|NonEmpty|, in which case these fields have type \li|dynamic|. When the
status of a bag~\li|b| changes from empty to non-empty
(lines~\ref{line:empty-to-nonempty:begin}--\ref{line:empty-to-nonempty:end})
or vice-versa
(lines~\ref{line:nonempty-to-empty:begin}--\ref{line:nonempty-to-empty:end}),
we reflect this change by updating the tag and the fields of~\li|b|.
At line~\ref{line:empty-to-nonempty:begin}, for instance, the permission that
describes~\li|b| is:
\begin{lstlisting}
  b @ Empty { head, tail : () }
      adopts cell a
\end{lstlisting}
After the tag update instruction, it is replaced with:
\begin{lstlisting}
  b @ NonEmpty { head, tail: () }
      adopts cell a
\end{lstlisting}
This structural permission may seem disturbing, because it cannot be
folded back to \qli|b @ bag a|. This is not a problem: at this point,
nothing requires us to produce the permission \qli|b @ bag a|. After
the second assignment, the current permission is:
\begin{lstlisting}
  b @ NonEmpty { head = c; tail: () }
      adopts cell a
\end{lstlisting}
Finally, after the last assignment, the current permission is:
\begin{lstlisting}
  b @ NonEmpty { head = c; tail = c }
      adopts cell a
\end{lstlisting}
Because we also have \qli|c @ dynamic| and because \li|dynamic| is duplicable,
this permission can be folded back to \qli|b @ bag a|, so that, when
\li|insert| completes, we are able to return \qli|b @ bag a|, as promised.

\end{longue}

\section{Other means of permitting sharing}
\label{sec:escape}

Adoption and abandon is not the only way of sharing mutable data.
We now describe two other mechanisms, namely nesting and
locks.

\subsection{Nesting}

\begin{longue}
\begin{figure}
\begin{lstlisting}
abstract nests (x : term) (p : perm) : perm
fact duplicable (nests x p)

val nest:
  [p : perm, a]
  exclusive a =>
  (x: a | consumes p) ->
  (| nests x p)

abstract punched (a : type) (p : perm) : type

val focus:
  [p : perm, a]
  exclusive a =>
  (consumes x: a | nests x p) ->
  (| x @ punched a p * p)

val defocus:
  [p : perm, a]
  (consumes (x: punched a p | p)) ->
  (| x @ a)
\end{lstlisting}
\caption{A simplified axiomatization of nesting}
\label{fig:nesting}
\end{figure}

\begin{figure}
\begin{lstlisting}
abstract lock (p: perm)
fact duplicable (lock p)
val new:  [p: perm] (| consumes p) -> lock p
val acquire: [p: perm] (l: lock p) -> (| p)
val release: [p: perm] (l: lock p
                     | consumes p) -> ()
\end{lstlisting}
\caption{A simplified axiomatization of locks}
\label{fig:locks}
\end{figure}

\end{longue}

Nesting~\cite{boyland-nesting-10} is a mechanism by which an object~$x$ adopts
(so to speak) a permission~$P$. It is a purely static mechanism. The act of
nesting $P$ in $x$ has no runtime effect, but consumes $P$ and produces a
witness, a permission which Boyland writes $\tynest Px$. Because nesting is
irreversible, such a witness is duplicable.

Once $P$ has been nested in $x$, whoever has exclusive ownership of~$x$ may
decide to temporarily recover~$P$. This is done via two symmetric operations,
say ``focus'' and ``defocus'', which in the presence of $\tynest Px$ convert between
$\tyatomic{x}{t}$ and $P \ast (P -\kern-2mm+ \tyatomic{x}{t})$ (where the
type~$t$ is arbitrary, but must be exclusive). The permission $P -\kern-2mm+
\tyatomic{x}{t}$ means that $P$ has been ``carved out'' of $x$.  While this is
the case, $\tyatomic xt$ is temporarily lost: in order to recover it, one must
give up $P$. Thus, it is impossible to simultaneously carve \emph{two}
permissions out of $x$.

Nesting subsumes F\"ahndrich and DeLine's adoption and
focus \cite{faehndrich-deline-02}.
We view it as a purely static cousin of adoption and abandon. Adoption is
more flexible in several important ways: it allows accessing two adoptees at
the same time, and allows abandoning an object forever. Nesting has
advantages over adoption and abandon: it cannot fail at runtime; it has no
time or space overhead; one may nest a permission, whereas one adopts an
object; and nesting is heterogeneous, i.e., an object~$x$ can nest multiple
distinct permissions, whereas, in the case of adoption and abandon, all
adoptees of $x$ must have the same type.

Nesting can be axiomatized in \mezzo as a trusted library, whose interface appears in%
\begin{courte}
the extended version of this paper~\cite{self-long}.
\end{courte}
\begin{longue}
\fref{fig:nesting}.
\end{longue}
In principle, this requires extending the proof of type
soundness; we have not done so. When applicable, nesting seems
preferable to adoption; however, adoption is more widely
applicable.

\begin{longue}
In the case of bags (\sref{sec:adoption}), \li|retrieve| takes a cell out of
the group in order to extract the element that it contains. If one chooses to
use nesting instead of adoption and abandon, then one cannot permanently take
the cell out of the group. Thus, the cell must remain in the group; but, in
that case, one must take the ownership of the element away from the cell. This
forces one to allow a cell to possibly contain no element, hence re-introduces
the need for a dynamic check.
\end{longue}

\subsection{Locks}
\label{sec:locks}

Dynamically-allocated locks in the style of concurrent separation
logic~\cite{ohearn-07,gotsman-storable-07,hobor-oracle-08,buisse-11} are
another dynamic mechanism for mediating access to a permission. A new lock, of
type $\tylock P$, where $P$ is an arbitrary permission, is created via a
function \li|new|. The functions \li|acquire| and \li|release| both take the
lock as an argument; \li|acquire| produces the permission~$P$, which
\li|release| consumes. The type $\tylock P$ is duplicable, so an arbitrary
number of threads can share the lock and simultaneously attempt to acquire it.
Within a critical section, delimited by \li|acquire| and \li|release|, the
``lock invariant''~$P$ is available, whereas, outside of it, it is not. The
``invariant''~$P$ can in fact be broken within the critical section, provided
it is restored when one reaches the end of the section.

Locks introduce a form of \emph{hidden state} into the language. Because the
permission $\tyatomic{l}{\tylock P}$ is duplicable, it can be captured by a
closure. As a result, it becomes possible for a function to perform a side
effect, even though its type does not reveal this fact (the pre- and postcondition
are empty). \mezzo's modest library for memoization exploits this feature.

Locks can be used to encode ``weak'' (duplicable) references in the style of
ML and duplicable references with affine content in the style of
Alms~\cite{tov-pucella-11}, both of which support arbitrary sharing.

Locks can be axiomatized in \mezzo as a library, whose interface appears in
\begin{courte}
the extended version of this paper~\cite{self-long}.
\end{courte}
\begin{longue}
\fref{fig:locks}.
\end{longue}
Again, the proof of type soundness must be extended; we have
begun this work. We view locks as complementary to adoption and abandon and
nesting. In a typical usage scenario, a lock protects an adopter, which in
turn controls a group of adoptees (or of nested permissions). This allows
a group of objects to be collectively protected by a single lock. It should
be noted that (we believe) adoption and abandon are sound in a concurrent
setting.

\section{Related work}
\label{sec:related}

The literature offers a wealth of type systems and program logics that are
intended to help write correct programs in the presence of mutable,
heap-allocated state. We review a few of them and contrast them with
\mezzo.

Ownership Types~\cite{clarke-potter-noble-98} and its descendants restrict
aliasing. Every object is owned by at most one other object, and an
``owner-as-dominator'' principle is enforced: every path from a root to an
object $x$ must go through $x$'s owner.
Universe Types~\cite{dietl-mueller-05} impose a
slightly different principle, ``owner-as-modifier''. Arbitrary paths
are allowed to exist in the heap, but only those that go through $x$'s owner
can be used to modify~$x$. This approach is meant to support program
verification, as it allows the owner to impose an object invariant.
Permission systems~\cite{bierhoff-aldrich-07,boyland-nesting-10,gordon-12} annotate
pointers not with owners, but with permissions. The permission carried by a
pointer tells how this pointer may be used (e.g. for reading and writing, only
for reading, or not at all) and how other pointers to the same object (if they
exist) might be used by others.

The systems mentioned so far are refinements (restrictions) of a traditional
type discipline. Separation logic~\cite{reynolds-02} departs from this
approach and obeys a principle that we dub ``owner-as-asserter''. (In
O'Hearn's words, ``ownership is in the eye of the
asserter''~\cite{ohearn-07}.) Objects are described by logical
assertions. To assert is to own: if one knows that ``$x$ is a linked list'',
then one may read and write the cells that form this list, and nobody else
may. Whereas the previously mentioned systems combine structural descriptions
(i.e., types) with owner or permission annotations, separation logic
assertions are at once structural descriptions and claims of ownership.

\mezzo follows the ``owner-as-asserter'' principle. In the future, this should
allow us to annotate permissions with logical assertions and use that as a
basis for the specification and proof of \mezzo programs. A tempting research
direction is to translate \mezzo into F$^\star$~\cite{fstar}. This purely
functional programming language is equipped with affine values, with powerful
facilities for expressing program specifications and proofs, and with a notion
of proof erasure.

Although our permission discipline is partly inspired by separation
logic~\cite{reynolds-02}, it is original in several ways. It presents itself
as a type system, as opposed to a program logic. This makes it less expressive
than a program logic, but more pervasive, in the sense that it can (and must)
be used at every stage of a program's development, without proof obligations.
It distinguishes between immutable and mutable data, supports first-class
functions, and takes advantage of algebraic data types in novel ways.

As far as we know, Ownership or Universe Types cannot express uniqueness or
ownership transfer. M\"uller and Rudich~\cite{mueller-rudich-07} extend
Universe Types with these notions. They rely on the fact that each object
maintains, at runtime, a pointer to its owner. The potential analogy with our
\texttt{adopter} fields deserves further study.

The use of singleton types to keep track of equations, and
the idea that pointers can be copied, whereas permissions
are affine, are inspired by Alias Types~\cite{alias-types-00}.
Linear~\cite{morrisett-al-07} and affine~\cite{tov-pucella-11} type systems
support strong updates and often view permissions (or ``capabilities'') as
ordinary values, which hopefully the compiler can erase. By offering an
explicit distinction between permissions and values, we guarantee that
permissions are erased, and we are able to make the flow of permissions mostly
implicit. Through algebraic data types and through the type constructor
$\tybar tP$, we retain the ability to tie a permission to a value, if
desired.

Regions~\cite{alias-types-00,faehndrich-deline-02,morrisett-al-07} have been
widely used as a technical device that allows a type to indirectly refer to a
value or set of values. In \mezzo, types refer to values directly. This
simplifies the meta-theory and the programmer's view.

Gordon \etal.~\cite{gordon-12} ensure data-race freedom in an extension
of C\#. They qualify types with permissions in the set \kw{immutable},
\kw{isolated}, \kw{writable}, or \kw{readable}. The first two roughly
correspond to our immutable and mutable modes, whereas the last two have no
\mezzo analogue. Shared (\kw{writable}) references allow legacy sequential
code to be considered well-typed. A salient feature is the absence of an alias
analysis, which simplifies the system considerably. This comes at a cost in
expressiveness: mutable global variables, as well as shared objects protected
by locks, are disallowed.

Plaid~\cite{aldrich-typestate-09} and \mezzo exhibit several common traits. A
Plaid object does not belong to a fixed class, but can move from one
``state'' to another: this is related to \mezzo's tag update. Methods carry
state pre- and postconditions, which are enforced via
permissions~\cite{bierhoff-aldrich-07}. Plaid is more ambitious in that states
are organized in an extensible hierarchy, whereas algebraic data types are
flat and closed.

\section{Conclusion and future work}
\label{sec:conclusion}

\mezzo is a high-level functional and imperative programming language where
the traditional concept of ``type'' is replaced with a more powerful concept
of ``permission''. Distinguishing duplicable, exclusive, and affine
permissions allows reasoning about state changes. We strive to achieve a
balance between simplicity and expressiveness by marrying a static discipline
of permissions and a novel dynamic form of adoption and abandon. By adding
other mechanisms for controlling sharing, such as nesting and locks, we
augment the expressiveness of the language and emphasize that the permission
discipline is sufficiently powerful to express these notions.
\mezzo is type-safe: well-typed programs cannot go wrong (but an abandon
operation can fail). We have carried out a machine-checked proof of type
safety~\cite{mezzo-proof}.
In the future, we would like to extend \mezzo with support for shared-memory
concurrency. We believe that, beyond locks (\sref{sec:locks}), many
abstractions (threads, channels, tasks, etc.) can be axiomatized so as to
guarantee that well-typed code is data-race-free.


\begin{longue}
  \appendix
  \section{Surface syntax}

So far, we have been fairly imprecise regarding the syntax of \mezzo. The
examples shown in sections \ref{sec:example} and \ref{sec:adoption} use special
syntactic conventions, while the formal definition of the typing rules
(\sref{sec:typing}) does not take them into account. This section clarifies the
rules for the syntax of \mezzo.

We saw in various examples that the user is
allowed to write \emph{name introductions}, \li|consumes| keywords, and benefits
from a special convention for function types; conversely, the internal syntax
has no such convention for function types, and makes no use of the other two
constructs.
Thus, there is a difference between the syntax that the user manipulates, which
we call the \emph{external syntax} (or \emph{surface} syntax), and the
representation that \mezzo internally uses, which we call the \emph{internal
syntax}.

Section \ref{sec:surface-syntax} briefly detailed the differences between the two
syntaxes, as well as the procedure one can use to convert from the former to the
latter.
We now formally define the syntax of \mezzo. We separate external constructs
from internal constructs; we introduce kind-checking rules; we define what it
means for a type in the surface syntax to be well-formed; we show how to
translate the constructs from the surface syntax into constructs from the
internal syntax. We also show that the translation preserves the well-kindedness
properties of a type.

\subsection{Differentiating the two syntaxes}

  \begin{figure}[t]
    \begin{center}
      \begin{tabularx}{\columnwidth}{rX}
        $T, t, P \hfill ::=$ & \hfill type or permission \\
        & \ldots  \\
        & $\tyarrow{t}{t}                   $ \hfill \emph{internal} function type \\
        & $\tyearrow{t}{t}                   $ \hfill \underline{\smash{\emph{external} function type}} \\
        & \ldots \\
        \\
        $e \hfill ::=$ & \hfill expression \\
        & \ldots  \\
        & $\elambda{x}{t}{t}{e}$ \hfill \emph{internal} anon. function \\
        & $\eLambda{X}{\kappa}{e}$ \hfill type abstraction \\
        & $\efun{\tyforalln{\vec{X}: \vec\kappa}{t}}{t}{e}$ \hfill \underline{\emph{external} anon. function} \\
        & \ldots 
      \end{tabularx}
    \end{center}
    \caption{Separating external and internal syntaxes}
    \label{fig:mezzo-types-ex}
  \end{figure}

  As section \ref{sec:surface-syntax} explained, the interpretation of an arrow
  type differs, depending on whether one uses the surface syntax or the internal
  syntax. To account for that difference, we \emph{clarify} what we mean when
  writing an arrow type: an arrow type is either the \emph{external variant}, or
  the \emph{internal variant}. These two arrows are now denoted by different
  symbols (\fref{fig:mezzo-types-ex}).

  Therefore, we do not see the external and internal syntaxes as two separate
  syntactical categories; rather, they are both restrictions of the general
  syntax of types. The external arrow, the name introduction, and the
  \li|consumes| keyword (all \underline{underlined}) may only appear in the
  external syntax, while the internal arrow may only appear in the internal
  syntax. As the name implies, the internal arrow is not exposed to the user.
  All other constructs may be used freely both in the external and internal
  syntaxes.

  Thus, translating from the external syntax to the internal version amounts to
  removing all the underlined constructs, replacing them with other constructs.

  Function expressions need to be translated as well, as an expression of the
  form $\efun{t_1}{t_2}{e}$ contains an implicit function type
  $\tyearrow{t_1}{t_2}$, meaning the function declaration benefits from the same
  syntactical conventions as the corresponding external function type.
  Furthermore, the external syntax for anonymous functions admits a \emph{type}
  as its argument, which is then interpreted as a pattern. For instance,
  $(\tynameintro xt, \tynameintro yu)$ is a tuple \emph{type}, which we
  interpret later on as a \emph{pattern} binding names $x$ and $y$. In order to
  clarify this as well, we introduce a syntax for \emph{internal} anonymous
  functions (\fref{fig:mezzo-types-ex}), which is closer to the one found in the
  $\lambda$-calculus. We write an internal function as
  $\elambda{x}{t_1}{t_2}{e}$, where $x$ is the name of the argument, bound in
  the function body $e$, $t_1$ the type of argument $x$ and $t_2$ the return
  type for the function.

\subsection{Binding rules}

\subsubsection{Environments}

  To keep track of the names that are available in the current context, we
  introduce naming environments.
  They are denoted as $\Gamma$, and are made up of lists of pairs $(x,
  \kappa)$ where $x$ is the name of the binder and $\kappa$ its kind.
  Adding a variable into an environment masks the previous definition of the
  variable. Concatenating environments with masking is done using a semicolon,
  as in $\Gamma_1; \dots ;\Gamma_n$.

  When merging several environments, we may wish to assert that the environments
  $\Gamma_i$ do not bind the same names. We use $\biguplus\Gamma_i$ for that
  purpose.

\subsubsection{Non-lexical scope}
\label{sec:kinding-appendix}

  \begin{figure*}
    \begin{center}
      \begin{mathpar}

        \inferrule[K-Var]
        {(x,\kappa) \in \Gamma}
        {\Gamma , s \vdash x: \kappa}

        \inferrule[K-Dynamic]
        {\quad}
        {\Gamma , s \vdash \tydynamic: \ktype}

        \inferrule[K-Arrow]
        {\Gamma , \kright \vdash t_1: \ktype\\
         \Gamma , \kright \vdash t_2: \ktype}
        {\Gamma , s \vdash \tyarrow{t_1}{t_2}: \ktype}

        \inferrule[K-EArrow]
        {\Gamma' = \Gamma; BV(t_1) \\
         \Gamma', \kleft \vdash t_1: \ktype\\
         \Gamma' \vdashb t_2: \ktype}
        {\Gamma , s \vdash \tyearrow{t_1}{t_2}: \ktype}

        \inferrule[K-Tuple]
        {\Gamma , s \vdash t_i: \ktype}
        {\Gamma , s \vdash \tytuple{t_1, \dots, t_n}: \ktype}

        \inferrule[K-Concrete]
        {\Gamma , s \vdash t_i: \ktype}
        {\Gamma , s \vdash \tyconcrete{A}{f_i: t_i}: \ktype}

        \inferrule[K-Forall]
        {\Gamma' = \Gamma; (x, \kappa') \\
        \Gamma' \vdashb t: \kappa}
        {\Gamma , s \vdash \tyforall{x}{\kappa'}{t}: \kappa}

        \inferrule[K-Exists]
        {\Gamma' = \Gamma; (x, \kappa') \\
        \Gamma' \vdashb t: \kappa}
        {\Gamma , s \vdash \tyexists{x}{\kappa'}{t}: \kappa}

        \inferrule[K-App]
        {\Gamma \vdashb t : \kappa_i \rightarrow \kappa \\
         \Gamma \vdashb t_i: \kappa_i}
        {\Gamma , s \vdash \tyapp{t}{t_i}: \kappa}

        \inferrule[K-NameIntro]
        {(x, \kterm) \in \Gamma\\
        \Gamma , s \vdash t: \ktype}
         {\Gamma , s \vdash (\tynameintro{x}{t}): \ktype}

        \inferrule[K-Consumes]
        {\Gamma , \kright \vdash t: \kappa \\
        \kappa = \ktype \text{ or } \kappa = \kperm}
        {\Gamma , \kleft \vdash \tyconsumes{t}: \kappa}

        \inferrule[K-Bar]
        {\Gamma , s \vdash t: \ktype \\
         \Gamma , s \vdash p: \kperm}
         {\Gamma , s \vdash \tybar{t}{p}: \ktype}

        \inferrule[K-Singleton]
        {(x, \kterm) \in \Gamma}
        {\Gamma , s \vdash \tysingleton{x}: \ktype}

        \inferrule[K-And]
        {\Gamma , s \vdash t_1: \kperm \text{ or } \Gamma , s \vdash t_1: \ktype\\
         \Gamma, s \vdash t_2: \kappa}
         {\Gamma , s \vdash \tyand{\mode{t_1}}{t_2}: \kappa}

        \inferrule[K-Empty]
          {\quad}
          {\Gamma , s \vdash \tyempty: \kperm}

        \inferrule[K-Perm]
          {\Gamma , s \vdash p: \kperm \\
          \Gamma , s \vdash q: \kperm}
          {\Gamma , s \vdash \tystar{p}{q}: \kperm}

        \inferrule[K-Anchored]
          {(x, \kterm) \in \Gamma\\
           \Gamma , s \vdashb t: \ktype}
          {\Gamma , s \vdash \tyatomic{x}{t}: \kperm}

        \inferrule[K-Extend]
          {\Gamma' = \Gamma; BV(t)\\
          \Gamma', \kright \vdash t: \kappa}
          {\Gamma \vdashb t: \kappa}
      \end{mathpar}
    \end{center}
    \caption{Kinding rules}
    \label{fig:kinding}
  \end{figure*}

  We want to define the kinding rules for the entire syntax of \mezzo; that is,
  we do not want a separate set of rules for both the external and the internal
  syntax. Thus, the kinding rules should operate on both syntaxes; in
  particular, we need to define kinding rules on the external syntax. This means
  that we ought to clarify the binding rules that govern the usage of our
  special name introduction construct $\tynameintro{x}{t}$.

  Our rules for binding names are non-standard, in the sense that we separate
  the \emph{name introduction} and the \emph{binding point}: our binders are
  \emph{not} lexically-scoped.

  As an example, consider the type of the \li|:=| function, which assigns a value
  into a reference (as seen in \sref{sec:surface-syntax}). Our references have a
  unique owner, which makes it possible for this function to change the type of
  the reference. As a consequence, the post-condition of the function needs to
  mention the function argument.
    $$
    \forall a,b.
    \tyearrow
    {(\tyconsumes{\tynameintro{x}{\kw{ref}\ a}}, \tyconsumes{b})}
    {\tybar{}{\tyatomic{x}{\kw{ref}\ b}}}
    $$
  In this arrow type, the name $x$ is made available both on the
  left-hand side of the function, and in the right-hand side; $x$ is not
  lexically scoped. We say that the left-hand side of this function type
  \emph{introduces} the name $x$, and that the binding point for the name $x$ is
  \emph{immediately above} the function type.

  The user may want to introduce names at arbitrary depth: under a \li|consumes|
  keyword, in the field of a concrete data type, in a tuple\ldots Therefore,
  even though a name may be introduced in-depth, we want it to be reachable as
  broadly as possible. For that purpose, we ``collect'' the names that appear in
  a type, and make them available in the entire type. In the example above,
  the name $x$ is collected and made available everywhere in the function type.

  The collection procedure should only descend into certain types, which we call
  \emph{transparent}. Transparent types are conjunctive in nature: structural
  types (tuples, concrete types) and conjunctions (mode and type, permission and
  type, permission and permission) are transparent.  It is unclear whether some
  constructs should be transparent or not; there is a design space to explore,
  and we chose to make some constructs \emph{opaque}.  For instance, names are
  not collected below quantifiers.

  We introduce the 
  $BV$ function (\fref{fig:bv}), where $BV(t)$ is the set of names
  introduced by type $t$.
  This function recursively descends into transparent constructs and stops at
  opaque constructs.
  Therefore, any traversal of a type will need to
  operate consistently with $BV$. More precisely, after traversing an opaque
  construct, it is mandatory to extend the working environment with the names
  found below the construct, before resuming the traversal.

  The kinding rules naturally operate consistently with $BV$, through the use of
  $\Gamma \vdashb t: \kappa$, which first extends $\Gamma$
  with the names found in $t$ before kind-checking $t$. As an example, the
  kind-checking rules for the opaque construct $\tyforall x \kappa u$ rely on
  $\vdashb$ in their premise.
  In the particular case of the external function type $\tyearrow{t_1}{t_2}$,
  the names that $t_1$ introduces are made available in $t_2$ as well.

  In the rest of the discussion, we assume environments to also contain the names
  of all the data types, along with their respective kinds
  (\fref{fig:bv}).

\subsection{Kinding rules}

Kind-checking is defined in \fref{fig:kinding}. The relation performs several
checks simultaneously:
\begin{itemize}
  \item it ensures that all types are well-kinded;
  \item it ensures that all names are bound according to the binding rules of
    the surface syntax;
  \item it ensures that $\tyconsumes$ annotations only appear in the left-hand
    side of an arrow type;
  \item it ensures that all type applications are complete; no partial type
    applications are allowed, as our system explicitly disallows it.
\end{itemize}

Our well-kindedness judgements are of the form $\Gamma \vdash t: \kappa$,
meaning that in environment $\Gamma$, the type $t$ satisfies all the above
conditions, and has kind $\kappa$. We intentionally restricted the form of rules
\Rule{K-Anchored} and \Rule{K-Singleton}; since the only types with kind
$\kterm$ are variables, we made it explicit that the only well-kinded singleton
types and atomic permissions are of the form $\tysingleton{x}$ and
$\tyatomic{x}{t}$.

In order to ensure that \li|consumes| annotations only appear in the left-hand
side of an arrow type, we introduce an extra $s$ variable. In the case of the
\Rule{K-Consumes}, the side $s$ has to be $\kleft$. Initially, the side is
$\kright$, and changes to $\kleft$ as soon as one sees an external arrow
(\Rule{K-EArrow}).

We define a $\vdashb$ variant, which sets the side to
$\kright$, meaning that no \li|consumes| annotation may appear here, and
extends the environment with the names introduced by $t$. As explained before,
the kind-checking relation needs to be consistent with the $BV$ function.
Therefore, this variant is used whenever an opaque construct is crossed, such as
in the premise of \Rule{K-Forall}. The relation $\vdashb$ is also the entry
point of the kind-checking relation, which we use to check top-level data types
definitions, as well as types that appear in expressions, for instance in
function definitions or type annotations.

One important rule is $\Rule{K-Arrow}$. Both $t_1$ and $t_2$ are checked in an
environment $\Gamma'$ which contains the names introduced by $t_1$. This means
that the names introduced in the domain $t_1$ of the function are available in both
the domain $t_1$ and the codomain $t_2$, while names introduced in the codomain
$t_2$ remain available in the codomain only.

  \subsection{Translating the surface syntax}
\label{sec:syntax-appendix}

  \begin{figure*}
    \begin{center}
      \begin{mathpar}
        \inferrule[T-Var]
        {\quad}
        {x \vartriangleright x}

        \inferrule[T-Dynamic]
        {\quad}
        {\tydynamic \vartriangleright \tydynamic}

         \inferrule[T-Tuple]
         {t_i \vartriangleright t'_i}
         {\tytuple{t_1, \dots, t_n} \vartriangleright \tytuple{t'_1, \dots, t'_n}}

         \inferrule[T-Concrete]
         {t_i \vartriangleright t'_i}
         {\tyconcrete{A}{f_i: t_i} \vartriangleright \tyconcrete{A}{f_i: t'_i}}

        \inferrule[T-App]
        {t_i \blacktriangleright t'_i\\
          t \blacktriangleright t'
        }
        {\tyapp{t}{t_i} \vartriangleright \tyapp{t'}{t'_i}}

        \inferrule[T-NameIntro]
        {t \vartriangleright t'}
        {\tynameintro{x}{t} \vartriangleright
          \tybar{\tysingleton{x}}{\tyatomic{x}{t'}}}

        \inferrule[T-Consumes]
        {t \vartriangleright t'}
        {\tyconsumes{t} \vartriangleright \tyconsumes{t'}}

        \inferrule[T-Forall]
        {t \blacktriangleright t'}
        {\tyforall{x}{\kappa}{t} \vartriangleright \tyforall{x}{\kappa}{t'}}

        \inferrule[T-Exists]
        {t \blacktriangleright t'}
        {\tyexists{x}{\kappa}{t} \vartriangleright \tyexists{x}{\kappa}{t'}}

        \inferrule[T-Bar]
        {t \vartriangleright t'\\
         p \vartriangleright p'}
        {\tybar{t}{p} \vartriangleright \tybar{t'}{p'}}

        \inferrule[T-Singleton]
        {\quad}
        {\tysingleton{x} \vartriangleright \tysingleton{x}}

        \inferrule[T-And]
        {t_1 \vartriangleright t'_1\\
         t_2 \vartriangleright t'_2}
         {\tyand{\mode{t_1}}{t_2} \vartriangleright \tyand{\mode{t'_1}}{t'_2}}

         \inferrule[T-Empty]
         {\quad}
         {\tyempty \vartriangleright \tyempty}

         \inferrule[T-Star]
         {p \vartriangleright p' \\ q \vartriangleright q'}
         {\tystar{p}{q} \vartriangleright \tystar{p'}{q'}}

        \inferrule[T-Anchored]
        {t \blacktriangleright t'}
        {\tyatomic{x}{t} \vartriangleright \tyatomic{x}{t'}}

        \inferrule[T-Extend-Exists]
        {\Gamma = BV(t) \\ t \vartriangleright t'}
        {t \blacktriangleright \exists\,\Gamma\ t'}

        \inferrule[T-EArrow]
        {\Gamma_1 = BV(t_1)\\
         t_1 \vartriangleright t'_1\\
         t_2 \blacktriangleright t'_2\\\\
         t'_{1,l} = [\tau/\tyconsumes\tau]t'_1\\
         t'_{1,r} = [\top/\tyconsumes\tau]t'_1
        }
         {\tyearrow{t_1}{t_2} \vartriangleright \forall\,\Gamma_1\
           \tyforall{r}{\kterm}{
             \tyarrow{
               \tybar{
                 \tysingleton{r}
               }{
                 \tyatomic{r}{t'_{1,l}}
               }
             }{
               \tybar{
                 t'_2
               }{
                 \tyatomic{r}{t'_{1,r}}
               }
             }
           }
        }

      \end{mathpar}
    \end{center}
    \caption{Translating types}
    \label{fig:translate-types}
  \end{figure*}

    \begin{figure*}
    \begin{center}
      \begin{mathpar}
        \inferrule[T-Annot]
        {t \blacktriangleright t'\\
         e \vartriangleright e'}
        {(\tynameintro{e}{t}) \vartriangleright (\tynameintro{e'}{t'})}

        \inferrule[T-ETApply]
        {t \blacktriangleright t'\\
         e \vartriangleright e'}
        {(\etapply{e}{t: \kappa}) \vartriangleright (\etapply{e'}{t': \kappa})}

        \inferrule[T-Fun]
      {
        \tyearrow{t}{u} \blacktriangleright
        \tyforall{\vec X'}{\vec\kappa'}{\tyarrow{t'}{u'}}\\
        p = \kw{type2pattern}(t)\\
        e \vartriangleright e'
      }
      {
        \efun{\tyforalln{\vec{X}: \vec\kappa}{t}}{u}{e}
        \vartriangleright\\\\
        \eLambda{\vec X}{\vec \kappa}{
        \eLambda{\vec X'}{\vec \kappa'}{
          \elambda{x}{t'}{u'}{\elet{p}{x}{e'}}
        }
        }
      }
      \end{mathpar}
    \end{center}
    \caption{Translating expressions}
    \label{fig:translating-expressions}
  \end{figure*}

  \begin{figure}
    \begin{center}
      \begin{tabularx}{\columnwidth}{llX}
        $BV(t)                        $ & $=$ & $\hfill \Gamma$ \\
        \\
        $BV(\tynameintro{x}{t})       $ & $=$ & $\hfill (x, \kterm)$ \\
        $BV(\tytuple{\vec{t}\,})      $ & $=$ & $\hfill \biguplus BV(\vec{t})$ \\
        $BV(\tyconcreteadopts{A}{\vec{f}:\vec{t}}{u})$ & $=$ & $\hfill \biguplus BV(\vec{t})$ \\
        $BV(\tybar{t}{P})             $ & $=$ & $\hfill BV(t)$\\
        $BV(\tyconsumes{T})           $ & $=$ & $\hfill BV(T)$\\
        $BV(X)                        $ & $=$ & $\hfill \kw{nil}$ \\
        $BV(\tyarrow{t}{t})           $ & $=$ & $\hfill \kw{nil}$\\
        $BV(\tyearrow{t}{t})          $ & $=$ & $\hfill \kw{nil}$\\
        $BV(\tyapp{T}{\vec{T}})       $ & $=$ & $\hfill \kw{nil}$\\
        $BV(\tyforall{X}{\kappa}{T})  $ & $=$ & $\hfill \kw{nil}$\\
        $BV(\tyexists{X}{\kappa}{T})  $ & $=$ & $\hfill \kw{nil}$\\
        $BV(\tysingleton{x})          $ & $=$ & $\hfill \kw{nil}$\\
        $BV(\tydynamic)               $ & $=$ & $\hfill \kw{nil}$\\
        $BV(\tyatomic{x}{t})          $ & $=$ & $\hfill \kw{nil}$\\
        $BV(\tyempty)                 $ & $=$ & $\hfill \kw{nil}$\\
        $BV(\tystar{P}{P})            $ & $=$ & $\hfill \kw{nil}$\\
        \\
        $BV(\kw{mutable? }\kw{data } d\ (\vec\tyvar: \vec\kappa)) $ & $=$ & $\hfill (d, \vec\kappa \rightarrow \ktype)$ \\
        $BV(\kw{abstract } d\ (\vec\tyvar: \vec\kappa): \kappa_r) $ & $=$ & $\hfill (d, \vec\kappa \rightarrow \kappa_r)$
      \end{tabularx}
    \end{center}
    \caption{Collecting the names introduced by a type}
    \label{fig:bv}
  \end{figure}

  In order to translate from the surface syntax down to the internal syntax,
  three constructs must be removed: name introductions, $\kw{consumes}$
  annotations and external arrows. We describe a set of transformations that
  perform the translation from the surface syntax down to the internal syntax.

  The translation step assumes that all well-formedness checks (as
  described in \sref{sec:kinding-appendix}) have been performed.

\subsubsection{Removing name introductions}

  We first need to remove all name introductions, that is, constructs of the
  form $\tynameintro{x}{t}$. The binding checks have been performed in the
  kind-checking process already; in order to ensure that the translation is
  faithful to the binding rules implemented by the kind-checking process, we
  need to make sure that for every call to $\vdashb$ performed in the
  kind-checking rules, we introduce explicit binders, either existential or
  universal.

  We introduce our translation $\vartriangleright$, and a variant
  $\blacktriangleright$, defined in rule \Rule{T-Extend-Exists}, which
  introduces explicit existential binders to account for the names introduced by
  a type. For every use of $\vdashb$ in the kinding rules, the translation
  makes use of $\blacktriangleright$.

  \begin{lemma}[Well-kindedness preservation]
    If $t$ is well-kinded, i.e. $\Gamma \vdashb t: \kappa$, and $t$ translates
    to $t'$, i.e. $t \blacktriangleright t'$, then $t'$ is similarly
    well-kinded, i.e. $\Gamma \vdashb t': \kappa$.
  \end{lemma}

  The one explicit call to $BV$ in the kinding rules (\Rule{K-EArrow}) is
  reflected by call to $BV$ in the corresponding translation rule
  (\Rule{T-EArrow}). This defines names introduced by the domain $t_1$ of a
  function type $\tyearrow{t_1}{t_2}$ to be universally quantified binders,
  enclosing an internal arrow.

  The rule \Rule{T-NameIntro} describes what one means when using a name
  introduction: $\tynameintro{x}{t}$ represents both a pointer to an element
  named $x$, and a permission stating that $x$ has type $t$. As
  $\vartriangleright$ is consistent with $BV$, we can assume that the
  variable $x$ has been introduced via an explicit binder above us.

  \begin{lemma}[Name introduction removals]
    If $t \blacktriangleright t'$, then $t'$ contain no occurrences of the name
    introduction construct.
  \end{lemma}

  As an example, consider a type $(\tynameintro{x}{t},
  \tysingleton{x})$. This type annotation describes the type of a tuple whose
  two components are equal; we name them $x$, and $x$ has type $t$. This type
  will be translated using $\blacktriangleright$ into:
  \[\exists (x: \kterm). (\typackage{x}{t}, \tysingleton{x})\]

\subsubsection{Interpreting \li|consumes| annotations}

  The \li|consumes| keyword receives a special treatment. The kind-checking
  rules made sure \li|consumes| keywords only appear in the left-hand side of
  external arrows. Rule \Rule{T-EArrow} therefore defines how to interpret
  \li|consumes| annotations that appear in the domain of an external arrow. This
  translation step changes the meaning of function types; hence, it goes
  from the external arrow $\rightsquigarrow$ to the internal arrow
  $\rightarrow$.

  The \li|consumes| keyword is all about ownership: intuitively, the ownership
  of a function argument will be, by default, returned to the caller, except for
  the sub-parts of the arguments that are marked with a \li|consumes| keyword.

  Let us now describe the meaning of the rule in greater detail.
  We write $[t_1/t_2]t$ as ``substitute
  $t_1$ for $t_2$ in $t$''; $\top$ is understood to mean either $\tyunknown$
  (which has kind $\ktype$) or $\tyempty$ (which has kind $\kperm$) depending on
  which one is appropriate.
  We introduce a name for the argument, so that
  we can talk about it in the post-condition of the function. We then take a
  permission for the ``full'' argument, and return a modified permission for the
  argument, where all parts marked as being consumed have been ``carved out''.

  One important property is that $t'_1$ no longer contains name introductions,
  as they have been removed by the recursive call to $\vartriangleright$. This is
  important: name introductions only make sense with the semantics of the
  external arrow and have no meaning when used within an internal arrow.

  \begin{lemma}[Consumes and external arrow removal]
    If $t \blacktriangleright t'$, then $t'$ contain no occurrences of the
    \li|consumes| keyword and of the external arrow $\rightsquigarrow$.
  \end{lemma}

  The \li|consumes| keyword can appear at any depth in the type of the argument: a
  sophisticated function may wish, for instance, to consume the ownership of
  just a single field of a data structure.

  As a final example, consider the type of the \li|swap| function, that swaps the
  two components of a mutable pair.
  The type of this function can be written, with the surface syntax conventions,
  as \li![a,b] (consumes x: mpair a b) -> (| x @ mpair b a)!, which we
  believe is concise, yet intuitive notation for the following internal type:
  $$
  \forall a.
  \forall b.
  \forall (x: \kterm)
  \tyarrow
  {\tybar{\tysingleton{x}}{\tyatomic{x}{\tyapp{\kw{mpair}{\ a\ b}}}}}
  {\tybar{}{\tyatomic{x}{\tyapp{\kw{mpair}{\ b\ a}}}}}
  $$

\subsubsection{Translating expressions}

  Expressions may contain types. Types appear for instance in type applications,
  which are used for instantiating polymorphic calls; they also appear in type
  annotations. We extend $\vartriangleright$ and define a translation for these
  expressions. The rules, which are to be found in
  \fref{fig:translating-expressions}, simply perform a translation of the types
  using $\blacktriangleright$.

  A more complex rule is \Rule{T-Fun}, which translates an external anonymous
  function into an internal one. Let us review the various steps of this
  translation. The external anonymous function already contains universal,
  user-provided quantifiers: these are translated using $\Lambda$-abstractions.
  The external arrow type $\tyearrow{t}{u}$ is translated
  using a combination of universal quantifications and an internal arrow type:
  we insert another set of $\Lambda$-abstraction to account for the implicit
  universal quantifications at kind $\kterm$. Next, we need to interpret $t$ as
  a pattern $p$; we omit the details of this procedure. Finally, the internal
  $\lambda$-abstraction takes a single argument which we name $x$; we recover the names
  that the user provided by binding the pattern $p$ to variable $x$ in the body
  of the $\lambda$-abstraction.

  We trivially extend $\vartriangleright$ to be the identity for the other
  constructs in the syntax of expressions.

  \begin{figure*}
  \centering
  \begin{mathpar}
    \inferrule[D-Concrete]{
      \kw{A}\text{ is a data constructor of }d\\
      d\text{ not defined as mutable}\\
      \modeD{\vec t}
    }
    {
      \modeD{\tyconcrete{A}{\vec f: \vec t}}
    }

    \inferrule[D-App]{
      \kw{data } d\ (\vec\tyvar: \vec\kappa) = \tyconcretev{A}{\vec f: \vec t} \\
      \modeD{\tyconcretev{A}{\vec f: [\vec u/\vec \tyvar] \vec t}}
    }
    {
      \modeD{\tyapp{d}{\vec u}}
    }

    \inferrule[D-Arrow]{}{\modeD{\tyarrow{t}{u}}}

    \inferrule[D-Forall]{\modeD t}{\modeD{\tyforall \tyvar \kappa t}}

    \inferrule[D-Exists]{\modeD t}{\modeD{\tyexists \tyvar \kappa t}}
    
    \inferrule[D-Singleton]{}{\modeD{\tysingleton x}}

    \inferrule[D-Bar]{
      \modeD{t}\\
      \modeD{P}
    }
    {
      \modeD{\tybar tP}
    }

    \inferrule[D-Dynamic]{}{\modeD{\tydynamic}}

    \inferrule[D-Anchored]{\modeD t}{\modeD{\tyatomic xt}}

    \inferrule[D-Empty]{}{\modeD\tyempty}

    \inferrule[D-Star]{\modeD p\\\modeD q}{\modeD{\tystar pq}}
  \end{mathpar}
  \caption{Definition of the ``$\kmodeD$'' judgement}
  \label{fig:dup}
\end{figure*}

\begin{figure*}
  \centering
  \begin{mathpar}
    \inferrule[X-Def]{
      \kw{data mutable } d\ (\vec\tyvar: \vec\kappa) = \ldots
    }
    {
      \modeX{d}
    }

    \inferrule[X-Concrete]{
      \kw{A}\text{ is a data constructor of }d\\
      \modeX{d}
    }
    {
      \modeX{\tyconcrete{A}{\vec f: \vec t}}
    }

    \inferrule[X-App]{
      \modeX{d}
    }
    {
      \modeX{\tyapp{d}{\vec t}}
    }
  \end{mathpar}
  \caption{Definition of the ``$\kmodeX$'' judgement}
  \label{fig:excl}
\end{figure*}

\tikzstyle{nuage}=[cloud, draw,
        text=black,
        font=\bfseries\sffamily\footnotesize]

\tikzstyle{plain}=[rectangle, draw,
        text width=1cm, text=black,
        minimum height=.5cm,
        text badly centered,
        font=\bfseries\sffamily\footnotesize]

\tikzstyle{txt}=[
        text width=2cm, text=black,
        minimum size=2cm,
        minimum height=.5cm,
        text badly centered,
        font=\ttfamily\footnotesize]

\makeatletter
\@ifundefined{WhizzyTeX}{
\begin{figure}
  \begin{center}
    \begin{tikzpicture}[>=stealth, node distance=1.5cm]
      \node [plain] (aff) {Affine};
      \node [plain, below left of=aff, text width=1.5cm] (dup) {Duplicable};
      \node [plain, below right of=aff, text width=1.5cm] (excl) {Exclusive};
      \node [plain, below of=aff, node distance=2cm, text width=1.5cm] (bot) {$\bot$};

      \begin{scope}[semithick]
        \draw [->] (excl) -- (aff);
        \draw [->] (dup) -- (aff);
        \draw [->] (bot) -- (dup);
        \draw [->] (bot) -- (excl);
      \end{scope}
    \end{tikzpicture}
  \end{center}
  \caption{The hierarchy of \emph{modes}}
  \label{fig:modes}
\end{figure}
}{}
\makeatother

\section{Modes and facts}
\label{sec:modes-appendix}

We have mentioned previously (\sref{sec:modes}, \fref{fig:typing-rules})
predicates of the form ``is duplicable'' or ``is exclusive''; ``exclusive'' and
``duplicable'' are \emph{modes}, which we formally introduce, along with the
corresponding lattice.  We then provide a formal definition for the two
predicates. Algebraic data types enjoy more complex predicates, which we call
\emph{facts}. We define the syntax of facts, along with the lattice they belong
to, and give details for their computation.

\subsection{Modes}

\subsubsection{Definition of modes}

Modes are predicates over types which form a lattice shown in \fref{fig:modes}.
These predicates only make sense for types at kind $\kperm$ and $\ktype$.
Data types, such as $\kw{list}$, have an arrow kind and enjoy more complex
predicates called \emph{facts}, which we detail in the following sections.
Types at kind $\kterm$ have no mode.

The $\kmodeA$ mode is a strict superset of $\kmodeD$ and $\kmodeX$
(\sref{sec:modes}). No object in \mezzo can be both $\kmodeD$ and $\kmodeX$,
i.e. no object can have mode $\bot$; however, the bottom element is required for the
fact computation which we will see later on. The mode lattice is thus complete.

We now define the subset of types that enjoy these predicates, which we write
respectively ``$\modeD{t}$'' and ``$\modeX{t}$'' (\fref{fig:dup},
\fref{fig:excl}). The two judgements are mutually exclusive and a type that
satisfies neither is $\kmodeA$.
The ``$\modeD{t}$'' judgement is defined co-inductively. 
The ``$\modeX{t}$'' is only defined for a small subset of types, as the only
exclusive types are concrete types and type applications that belong to an
mutable-defined data type.

\begin{lemma}[Semantics of $\kmodeX$]
If $K \vdash t: \ktype$,
$t$ is exclusive $\Leftrightarrow
\tystar
  {\tyatomic xt}
  {\tyatomic yt} \Vdash x\text{ and }y\text{ are distinct values}
$
\end{lemma}

\begin{lemma}[Semantics of $\kmodeD$]
If $K \vdash p: \kperm$,
$p$ is duplicable $\Leftrightarrow
p \Vdash \tystar pp$.
If $K \vdash t: \ktype$, $t$ is duplicable if and only if $\tyatomic xt$ is
duplicable.
\end{lemma}

\subsubsection{Mode constraints}

A function may wish to request that a type be duplicable. Consider the
(incorrect) signature one may write for the $\kw{find}$ function:
\begin{lstlisting}
  val find: [a] (list a, a -> bool) -> option a
\end{lstlisting}
The function returns the first element in a list that satisfies the given
predicate, if any. However, this function will only type-check if ``$a$'' is
specialized to a duplicable type, as the function will alias, in the case of
success, the element found in the list, by returning a pointer to it.

A better solution is for the function to demand that its type parameter ``$a$''
be duplicable. We extend the syntax of types (\fref{fig:mezzo-types-and}) with
\emph{mode constraints}, thus allowing us to write the following type:
\begin{lstlisting}
  val find: [a]
    (duplicable a | list a, a -> bool) -> option a
\end{lstlisting}
When mode constraints are used for a function's argument, we provide syntactic
sugar, for readability.
\begin{lstlisting}
  val find: [a] duplicable a =>
    (list a, a -> bool) -> option a
\end{lstlisting}

  \begin{figure}[t]
    \begin{center}
      \begin{tabularx}{\columnwidth}{rX}
        $T, t, P \hfill ::=$ & \hfill type or permission \\
        & \ldots  \\
        & $\tyand{m\ \alpha}{t} $ \hfill \underline{\smash{mode constraint and type conjunction}} \\
        & \ldots
      \end{tabularx}
    \end{center}
    \caption{Extending the syntax with mode constraints}
    \label{fig:mezzo-types-and}
  \end{figure}

\subsection{Facts}

We claimed earlier (\sref{sec:modes}) that the programmer, and the type-checker,
can always tell what mode a permission satisfies. The coinductive definition of
the ``is duplicable'' judgement allows one to check that a given type satisfies
a mode, but provides no way to \emph{compute} ``the'' mode of a type.

One could read \fref{fig:dup} as a set of recursive rules that, when applied,
perform a computation. This would be sound, but too strict. Indeed, looking at
\Rule{D-App}, we see that a type application is duplicable if and only if all
its unfoldings are duplicable. For instance, $\tylist t$ is duplicable if both
$\kw{Nil}$ and $\tyconcrete{\kw{Cons}}{ \khead: t; \ktail: \tylist t }$ are
duplicable. A recursive reasoning, for lack of a co-inductive ``$\modeD{\tylist
t}$'' hypothesis, would determine $\tyconcrete{\kw{Cons}}{ \khead: t; \ktail:
\tylist t }$ to be affine and conclude that $\tylist t$ is affine as well.

One could also perform several attempts, first trying to show that ``$\tylist t$
is duplicable'' then, failing that, try to show that ``$\tylist t$ is
exclusive''. This would be terribly inefficient, because, not knowing the mode
of $t$, we would need to perform an (exponentially expensive) search on the
modes of any type applications that may be found in $t$.

\newcommand{\fact}[2]{%
  \kw{fact}(#1,#2)
}

The problem lies with algebraic data types, as the mode of a type application
depends on the parameters. We cannot afford to compute the mode of $\tylist t$
for every possible value of $t$. What we want instead is to state the following
\emph{fact}: ``if $t$ is duplicable, then $\tylist t$ is duplicable''.
We write $\fact {\tylist\alpha}{D\ \alpha \Rightarrow D}$, meaning
that $D\ \alpha \Rightarrow D$ is a valid fact for $\tylist\alpha$.

\subsubsection{Syntax of facts}

\newcommand{\bigWedge}{\ensuremath{\bigwedge\hspace{-.8em}\bigwedge}}
\newcommand{\sep}{\ensuremath{\;|\,}}

\begin{figure}
  \centering
  \begin{tabularx}{\columnwidth}{rX}
    $f ::= \forall\vec\alpha. \bigWedge \vec\imath$ & \hfill fact \\
      \\

    $i ::= h \Rightarrow c$ & \hfill implication \\
      \\

    $c ::= m$ & \hfill conclusion \\
      \\

    $m ::= A \sep X \sep D \sep \bot$ & \hfill mode \\
      \\

    $h ::= \kfalse \sep \bigWedge \vec m\ \vec\alpha$ & \hfill hypothesis \\

  \end{tabularx}
  \caption{Syntax of facts}
  \label{fig:syntax-facts}
\end{figure}

The syntax of facts is presented in \fref{fig:syntax-facts}. A fact $f$ for type
$t$ is a conjunction of implications. For each implication, if the hypotheses
are satisfied, then the conclusion is a valid mode for $t$. A trivial fact that
any type $t$ enjoys is: ``$\ktrue \Rightarrow A$''. The fact for
$\kw{list}$ that we just saw is another syntactically correct fact.

The syntax of facts defined in \fref{fig:syntax-facts} is specialized over a set
of parameters $\vec\alpha$. These are the formal parameters of the data type that
the fact refers to. For instance, when computing the fact for type
$\tylist\alpha$, there is only one parameter $\alpha$.

For a fixed set of parameters $\vec\alpha$, facts form a upward-closed
semi-lattice, whose $\top$ element is ``$\ktrue \Rightarrow A$''.

\subsubsection{Operations on the facts lattice}

Before jumping into the definition of fact inference, we need to equip our
lattice with several operations. Moreover, we assume facts in the lattice to be
total, that is, to have exactly four implications, one for each possible mode.
Ensuring a fact is total is easy: we just need to extend its conjunction with
implications of the form ``$\kfalse \Rightarrow \vec m$'' for the missing modes
$\vec m$.

We also assume our hypotheses to be total, that is, to either have exactly one
clause of the form $m\ \alpha$ for each $\alpha$, or to be $\kfalse$. We then
define $\ktrue$ to be a synonym for either $\bigWedge \bot\ \alpha$, or the
empty conjunction.

\begin{definition}[Constant fact]
  A constant mode $m$ maps into a fact as follows:
  \[
    \kw{constant}(m) =
    \begin{cases}
      \ktrue \Rightarrow m' & \text{ if } m' \geq m \\
      \kfalse \Rightarrow m' & \text{ otherwise } \\
    \end{cases}
  \]
\end{definition}

\begin{definition}[Conjunction of clauses]
  The conjunction of clauses $m\,\alpha$ and $m'\,\alpha$, denoted as
  $m\, \alpha \wedge m'\, \alpha$, is $(m \sqcap m')\,\alpha$, where $\sqcap$ is
  the natural ``meet'' operation induced by the lattice of modes.
\end{definition}

\begin{definition}[Conjunction of hypotheses]
  The conjunction of hypotheses $h \wedge h'$ is the pairwise conjunction of
  clauses whose parameter $\alpha$ match, or $\kfalse$ if either $h$ or $h'$ is
  false.
\end{definition}

\begin{definition}[Join operation]
  We define the join operation $\sqcup$ on our lattice as follows.
  \[
    (\vec h \Rightarrow \vec m) \sqcup (\vec {h'} \Rightarrow \vec m) =
    \overrightarrow{ h \wedge h'} \Rightarrow \vec m
  \]
\end{definition}

\begin{definition}[Fact for a parameter]
  If $\alpha$ is a parameter of the type whose fact is currently being inferred,
  then the fact for $\alpha$ is:
  \[
    \kw{parameter}(\alpha) = \bigWedge (m\ \alpha) \Rightarrow m
  \]
\end{definition}

We also need a meet operation for our lattice. Indeed, if we encounter the
conjunction of a mode hypothesis and a type, such as $\tyand {m\ \alpha}t$, the
fact environment $F$ need to be refined to take into account the new hypothesis
for $\alpha$. This means that both facts $F(\alpha)$ and $m\ \alpha$ will
hold when inferring the fact for $t$. The conjunction of two facts corresponds
to a meet operation on our lattice.

The ``meet'' of $f_1$ and $f_2$ is:
\[
  f_1 \sqcap f_2 = \overrightarrow{h_1 \vee h_2} \Rightarrow \vec m
\]
However, this general operation cannot be defined, as our hypotheses are made up
of conjunctions, and do not handle disjunction. We therefore define the ``meet''
operation on our lattice only in the case where one of the arguments is a
constant fact.

\begin{definition}[Meet operation]
  Assuming that $f_1$ is a constant fact, that is, that all hypotheses in $f_1$
  are either $\ktrue$ are $\kfalse$, we define the ``meet'' of $f_1$ and $f_2$
  to be:
  \[
    f_1 \sqcap f_2 = \bigWedge \begin{cases}
      h_2 \Rightarrow m \text{ if } \kfalse \Rightarrow m \in f_1\\
      \ktrue \Rightarrow m \text{ otherwise } (\text{meaning }\ktrue \Rightarrow m \in f_1)
    \end{cases}
  \]
\end{definition}

\subsubsection{Fact inference}

We introduce a procedure that infers the ``best'' fact $f$ for a type $t$, that
is, the smallest element $f$ in the lattice such that $\fact ft$.

\begin{lemma}[Existence and unicity]
  Given a type $t$, there exists a unique fact $f$ that satisfies $\fact ft$ as
  well as:
  $$\forall f'. \fact {f'}t \Rightarrow f \leq f'$$
\end{lemma}

Our inference judgements are of the form $F \vdash \fact tf$, meaning that if
$F$ contains facts for other data types, then the smallest fact we can infer for
$t$ is $f$.  For groups of mutually recursive data type definitions, the
type-checker of \mezzo leverages this procedure and determines the smallest fact
for each type, through the use of a fixed-point computation. 
Our fact environment $F$ maps a data type $t$ to a fact $f$ that is universally
quantified over the type's parameters. Namely, $F(t)$ is of the form
$\forall\vec\alpha. \fact {(\tyapp t{\vec\alpha)}}f$.

What does it mean for a fact environment $F$ to be correct? A fact environment
is correct if the following rule holds.
\begin{mathpar}
  \inferrule[Facts-Correct]{
    F' = F\big[\vec\alpha\mapsto\kw{parameter}(\vec\alpha)\big] \\
    \kw{definition}(\tyapp t{\vec\alpha}) = u\\
    F' \vdash \fact u{f_{\vec\alpha}}\\
    \forall\vec\alpha. f_{\vec\alpha} \leq F(t)
  }{
    F\text{ is correct}
  }
\end{mathpar}
We then take an initial assignment $F_0$ where each type is assigned the top
fact, and iterate the process until we reach a fixed point.

\begin{figure*}[t]
  \begin{mathpar}

    \inferrule[Fact-Dynamic]{}{
      F \vdash \fact \tydynamic{\kw{constant}(D)}
    }

    \inferrule[Fact-Singleton]{}{
      F \vdash \fact {\tysingleton x}{\kw{constant}(D)}
    }

    \inferrule[Fact-Arrow]{}{
      F \vdash \fact {\tyarrow tu}{\kw{constant}(D)}
    }

    \inferrule[Fact-Var]{
    }{
      F \vdash \fact \alpha{F(\alpha)}
    }

    \inferrule[Fact-App]{
      F(u) = f_0\\
      F \vdash \overrightarrow{\fact {v_i}{f_i}}\\
      f = \kw{compose}(f_0, \vec f)
    }{
      F \vdash \fact{\tyapp {u}{\vec v}}{f}
    }

    \inferrule[Fact-And]{
      F' = \kw{assume}(F, m\ \alpha)\\
      F' \vdash \fact tf
    }{
      F \vdash \fact {\tyand {m\ \alpha}t}f
    }

    \inferrule[Fact-Forall]{
      F' = \kw{assume}(F, \bot\ \alpha)\\
      F' \vdash \fact tf
    }{
      F \vdash \fact {\forall\alpha. t}f
    }

    \inferrule[Fact-Exists]{
      F' = \kw{assume}(A\ \alpha)\\
      F' \vdash \fact tf
    }{
      F \vdash \fact {\exists\alpha. t}f
    }

    \inferrule[Fact-Tuple]{
      F \vdash \fact {\vec t}{\ldots \wedge \vec h \Rightarrow D \wedge \ldots}
    }{
      F \vdash \fact {(t_1, \cdots, t_n)}{\bigWedge \vec h \Rightarrow D}
    }

    \inferrule[Fact-Concrete-X]{
      \kw{A}\text{ is a constructor of }t\\
      t\text{ is defined as mutable}
    }{
      F \vdash \fact {\tyconcrete{A}{\vec f: \vec t}}X
    }

    \inferrule[Fact-Concrete-D]{
      \kw{A}\text{ is a constructor of }t\\
      t\text{ is defined as immutable}\\\\
      F \vdash \fact {\vec t}{\ldots \wedge \vec h \Rightarrow D \wedge \ldots}
    }{
      F \vdash \fact {\tyconcrete{A}{\vec f: \vec t}}{\bigWedge \vec h \Rightarrow D}
    }

    \inferrule[Fact-Bar]{
      F \vdash \fact t{f_1}\\
      F \vdash \fact P{f_2}
    }{
      F \vdash \fact {\tybar tP}{f_1 \sqcup (f_2 \sqcap \kw{constant}(X))}
    }

    \inferrule[Fact-Empty]{}{
      F \vdash \fact \tyempty{\kw{constant}(D)}
    }

    \inferrule[Fact-Anchored]{
      F \vdash \fact t{h \Rightarrow X \wedge \vec\imath}
    }{
      F \vdash \fact {\tyatomic xt}{\vec\imath}
    }

    \inferrule[Fact-Star]{
      F \vdash \fact p{f_1}\\
      F \vdash \fact q{f_2}
    }{
      F \vdash \fact {\tystar pq}{f_1 \sqcup f_2}
    }
 
  \end{mathpar}
  \caption{Fact inference}
  \label{fig:fact-inference}
\end{figure*}

Let us briefly comment the fact inference rules, which are exposed in
\fref{fig:fact-inference}.
As stated in the \Rule{Facts-Correct} judgement, we infer a fact in an extended
environment $F'$, where the ``best'' fact is assigned to the type's parameters.
Therefore, rule \Rule{Fact-Var} consists in a mere lookup in the environment.
Rule \Rule{Fact-And} refines the current fact for a type variable using the
``meet'' operation on the lattice of facts.
Rule \Rule{Fact-Tuple} only has one non-trivial conclusion: a tuple is
duplicable if all its components are.
There are two rules for concrete data types. Rule \Rule{Fact-Concrete-X} tells
that a concrete data type that ``belongs'' to a mutable data type is exclusive.
Rule \Rule{Fact-Concrete-D} is for constructors that belong to an immutable data
type; this rule is analogous to \Rule{Fact-Tuple}.
Rule \Rule{Fact-Anchored} removes any implication whose conclusion is $X$, as
the ``$p$ is exclusive'' judgement has no meaning when $p$ is at kind $\kperm$.
Rule \Rule{Fact-Bar}, conversely, extends the fact for the permission with a
$X$ constant fact, so that the resulting conjunction of a type and a permission
can be seen as exclusive if the type itself is.

Rule \Rule{Fact-App} is slightly more involved, and requires us to explain the
$\kw{compose}$ function, which relates together the fact of the data type
currently examined, and the pre-computed fact for another data type.
Let us take an example before formally defining $\kw{compose}$. We consider the
following data type definition:
\[
  \kw{data } \kw{listpair} \ \alpha\ \beta =
  \tyconcrete{ListPair}{\kw{listpair}: \tylist{(\alpha,\beta)}}
\]
The initial fact for ``$\alpha$'' (resp. ``$\beta$'') is
``$\kw{parameter}(\alpha)$''
(resp. ``$\kw{parameter}(\beta)$''). Thus, the inferred fact for
``$(\alpha,\beta)$'' is ``$D\ \alpha \wedge D\ \beta \Rightarrow D$''. Besides,
assuming the environment $F$ contains the inferred fact for $\kw{list}$ already,
we have:
\[
  \forall \gamma. \fact {\tylist\gamma}{D\ \gamma \Rightarrow D}
\]
The value of the formal parameter ``$\gamma$'' for $\kw{list}$ is
``$(\alpha,\beta)$''.
Therefore, we need ``$(\alpha,\beta)$'' to be duplicable for
``$\tylist{(\alpha,\beta)}$'' to be duplicable. Knowing that:
\[
  \fact {(\alpha,\beta)}{D\ \alpha \wedge D\ \beta \Rightarrow D}
\]
we find that ``$D\ \alpha \wedge D\ \beta$'' is a sufficient condition for
``$\tylist{(\alpha,\beta)}$ is duplicable'' to hold. This is also a necessary condition.
Therefore, we infer that, inside the definition of $\kw{listpair}$, the fact for
``$\tylist{(\alpha,\beta)}$'' is ``$D\ \alpha \wedge D\ \beta \Rightarrow D$''.

\begin{definition}[Fact composition]
  We consider a type application ``$u\ \vec v_i$''.
  Assume the (generalized) inferred fact for type $u$ is ``$\forall
  \vec\alpha_i. f$'', and that facts $\vec f_i$ are the inferred facts
  for the effective parameters $\vec v_i$. The
  composition of $f$ with $\vec f_i$ is:
  \[
    \kw{compose}(f,\vec f_i) = f\big[
      \kw{hyp}(m, i) \big/ m\ \alpha_i
    \big]
  \]
  where $\kw{hyp}(m, i)$ finds the hypothesis associated to the conclusion $m$
  in fact $f_i$. Our facts are assumed to be total, so $\kw{hyp}(m, i)$ is
  always defined.
\end{definition}

\subsubsection{Abstract types, type abbreviations}

When sealing a module with a signature, and making a type abstract, some
properties about the type still need to be exposed in the signature. Variance is
an example of such a property that one can reveal in the signature of an OCaml
module.

If the $\kw{list}$ type were to be abstract, in the absence of any other
information about it, \mezzo would consider $\tylist t$ to be affine, for any
type $t$. We must therefore allow the user to specify facts about abstract types
in module signatures. The syntax is:
\begin{lstlisting}
  abstract list a
  fact duplicable a => duplicable list a
\end{lstlisting}
We check that the exposed fact in the signature is equal to the inferred one.

\mezzo also offers non-recursive type abbreviations, which the user can choose
to export as abstract types. The non-recursivity restriction allows the
type-checker to infer their facts using the same mechanism that we exposed.

\end{longue}

\bibliographystyle{plain}
\hbadness=10000 
\bibliography{english,local,self}

\end{document}